\documentclass{article}

\usepackage{arxiv}

\usepackage[utf8]{inputenc} 
\usepackage[T1]{fontenc}    
\usepackage{hyperref}       
\usepackage{url}            
\usepackage{booktabs}       
\usepackage{amsfonts}       
\usepackage{nicefrac}       
\usepackage{microtype}      
\usepackage{lipsum}
\usepackage{amsmath,amssymb,amsfonts, epsf}
\usepackage{breqn}
\usepackage{algorithm}
\usepackage{algorithmic}
\usepackage{mathtools}
\usepackage{graphicx}
\usepackage{cases}
\usepackage{color}
\usepackage{float}
\usepackage{amssymb}
\usepackage{appendix}
\graphicspath{{Fig/}}
\usepackage{nomencl}
\usepackage{mathtools}
\usepackage{booktabs}
\usepackage{nicefrac} 
\usepackage{microtype}      
\DeclareMathOperator*{\argmax}{arg\,max}
\usepackage{hyperref}
\hypersetup{
    colorlinks=true,
    linkcolor=blue,
    filecolor=magenta,      
    urlcolor=cyan,
}
\usepackage{xcolor}

\title{Real-Time Model Calibration with Deep Reinforcement Learning}

\author{Yuan Tian  \thanks{These authors contributed equally to this work.} \\
	ETH Zurich\\
    \texttt{yutain@ethz.ch} \\
    \And
    Manuel Arias Chao $^*$ \\
    ETH Zurich\\
   \texttt{manuel.arias@ethz.ch}\\
   \And
    Chetan Kulkarni\\
	KBR, Inc., NASA Ames Research Center\\
	\texttt{chetan.s.kulkarni@nasa.gov}\\
   \And
   Kai Goebel\\
   Lule{\aa} University of Technology\\
   \texttt{kai.goebel@ltu.se}\\
   \And	
   Olga Fink \\
   ETH Zurich\\
   \texttt{ofink@ethz.ch}\\
}

\begin{document}
\maketitle

\begin{abstract}
The dynamic, real-time, and accurate inference of model parameters from empirical data is of great importance in many scientific and engineering disciplines that use computational models (such as a digital twin) for the analysis and prediction of complex physical processes.
However, fast and accurate inference for processes with large and high dimensional datasets cannot easily be achieved with state-of-the-art methods under noisy real-world conditions. The primary reason is that the inference of model parameters with traditional techniques based on optimisation or sampling often suffers from computational and statistical challenges, resulting in a trade-off between accuracy and deployment time. In this paper, we propose a novel framework for inference of model parameters based on reinforcement learning. The contribution of the paper is twofold: 1) We reformulate the inference problem as a tracking problem with the objective of learning a policy that forces the response of the physics-based model to follow the observations; 2) We propose the constrained Lyapunov-based actor-critic (CLAC) algorithm to enable the robust and accurate inference of physics-based model parameters in real time under noisy real-world conditions. The proposed methodology is demonstrated and evaluated on two model-based diagnostics test cases utilizing two different physics-based models of turbofan engines. The performance of the methodology is compared to that of two alternative approaches: a state update method (unscented Kalman filter) and a supervised end-to-end mapping with deep neural networks. The experimental results demonstrate that the proposed methodology outperforms all other tested methods in terms of speed and robustness, with high inference accuracy.
\end{abstract}

\keywords{model calibration \and reinforcement learning \and model-based \and diagnostics \and deep learning}

\section{Introduction}

\begin{figure}[ht]
\centering
\includegraphics[width=8cm]{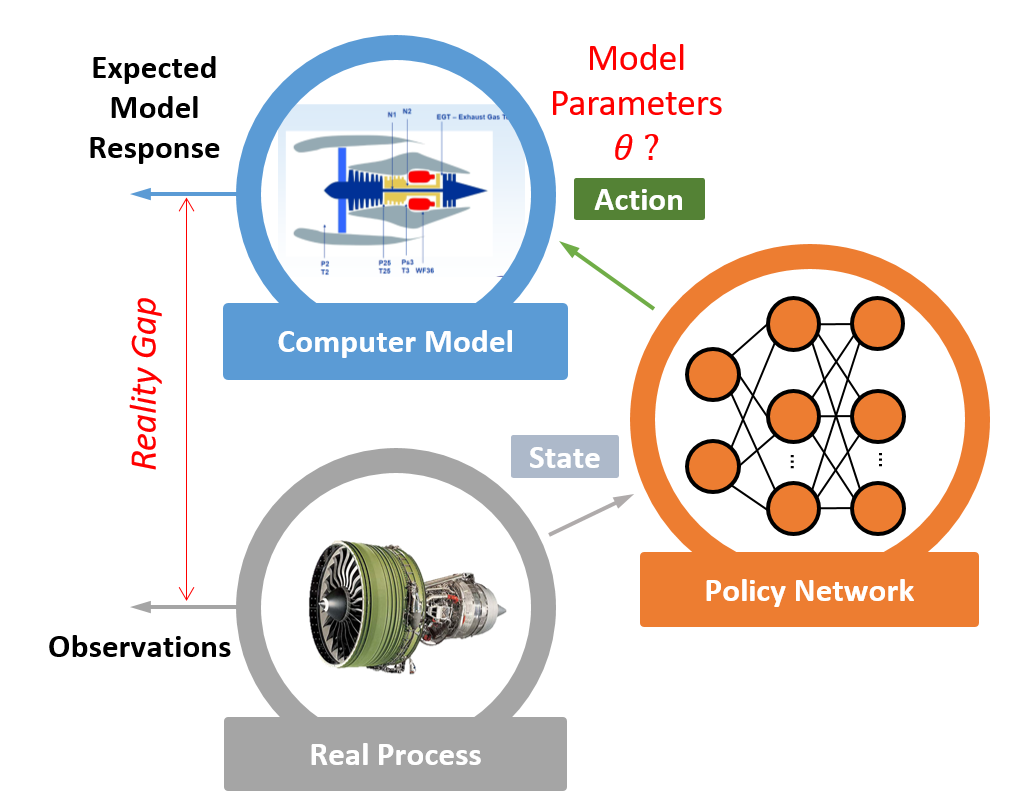}
\caption{\textbf{Calibration of physics-based models} aims to infer model parameters that make the physics-based model response follow the observations, thus reducing the \emph{reality gap}. In this work, a reinforcement learning algorithm is used to obtain a neural network policy that bridges the gap between physics-based model predictions and observations in real time.}
\label{fig:calibration}
\end{figure}

Inference of computational model parameters from empirical data can be referred to as \emph{model calibration} \cite{Kennedy2001}. Model calibration aims to both obtain model parameters that are theoretically plausible and generate model predictions that fit the observations. The inferred model parameters often represent physical quantities that are not directly observable or observed, i.e., they are not directly obtained from sensor measurements. Therefore, the inference of physics-based model parameters enables one to understand the underlying reasons for a discrepancy between physics-based model predictions and observations, i.e., the \emph{reality gap} (see Figure \ref{fig:calibration}). This is of particular relevance for scientific and engineering disciplines where one is interested in improving the physics-based model analytically or explaining the observed process in light of a given physics-based model structure. Applications can be found in multiple areas, including geology \cite{Elsheikh2015}, climatology \cite{Sanso2008}, biology \cite{Henderson2009}, health \cite{Rutter2009}, finance \cite{Liu2019, Deng2008}, cognitive science \cite{Kangasraasio2019}, mechanical engineering \cite{Kumar2013}, and applied physics \cite{Higdon2008}.

A particularly important field of application aiming at a reasoned analysis of discrepancies between model predictions and observations is \textbf{model-based system health diagnostics} of safety-critical engineered systems. Diagnostics involves detecting when a fault occurs, isolating the root cause, and identifying the extent of the damage \cite{Roychoudhury2013}. In model-based health diagnostics, the discrepancy between model and observation is interpreted as a deteriorated or anomalous response of the system. Therefore, model-based health diagnostics addresses the diagnostics problem by inferring the value of model parameters, representing the health condition of the sub-components of a system that make the physics-based model predictions fit the observations. In this way, anomalies in the system's behavior are detected and characterized by the value of model parameters.

Because of the relevance of model calibration in applications such as the one presented above, it is important that model calibration provide accurate inference of the model parameters while being robust to uncertainty in the observations and the physics-based model structure. However, calibration in real-world scenarios faces \textbf{computational and statistical difficulties}. Computational issues are related to the need for running time-consuming simulations using optimisation and inference techniques that generally imply a trade-off between inference accuracy and computation time. Scaling the method to \textbf{large datasets}, \textbf{high dimensional spaces} and \textbf{complex dynamic models} (such as a model with flow field calculation) further exacerbates the problem. Statistical issues arise from a) the incompleteness of the model representation, b) the existence of multiple solutions, i.e., \textit{confounding solutions} that match the observations, and c) the uncertainty of the observations. Some safety-critical applications, such as model-based diagnostics of aircraft engines, require simultaneous speed, accuracy, and robustness in the inference of the model parameters to enable a fast and reliable state assessment. The necessity of fulfilling all of these requirements at the same time makes the development of methods for reliable dynamical model calibration challenging. 

Several methods have been proposed to address the problem of dynamical model calibration. When the physics-based model structure is well founded on known physical principles (e.g., aircraft thermodynamic engine models), the majority of the available methods for parameter inference are \textbf{probabilistic or estimation approaches} developed in the fields of \textbf{optimal control} \cite{Crassidis2011} and \textbf{statistics} \cite{Sacks1989}. Some examples of popular estimation methods include iterative reweighted least squares schemes \cite{AriasChao2015}, Kalman filters (KF) \cite{Kalman1960}, extended Kalman filters (EKF) \cite{Einicke1999, Borguet2012}, unscented Kalman filters (UKF) \cite{Julier1997, Turner2010}, particle filters \cite{Kantas2015} or Bayesian inference methods using Markov chain Monte Carlo \cite{Rutter2009}. Approaches of this type scale relatively well to high-dimensional calibration problems and, with their probabilistic nature, handle observation noise reasonably well. These estimation methods have achieved good results in practical applications and are considered the state-of-the-art in several applications such as model-based diagnostics. Yet, despite these attractive properties, they all suffer, at least to some degree, from various computational and statistical difficulties in real-world scenarios. In particular, this is because estimation with these methods involves multiple evaluations of the computational model, which makes them unsuitable for real-time calibration of models based on large datasets when the available computational resources are limited. Moreover, these methods are particularly affected by the inadequacy of the physics-based model structure, resulting in an inaccurate characterization of the reality gap. 

More recently, \textbf{data-driven approaches} have been proposed to calibrate physics-based models. Aiming to avoid time-consuming simulations of previous calibration methods and achieve real-time model calibration, some researchers have deviated from the probabilistic formulation of the calibration problem. The most common approach is to address the calibration problem as a supervised learning problem \cite{Liu2019}. In this case, a neural network algorithm is trained in the inverse relation between the observations and the model parameters. Although these methods provide a real-time calibration approach (only a forward pass over a neural network is required at deployment time), the accuracy of the methods is strongly dependent on the representative quality of the training datasets. As a result, this model calibration approach is not able to adapt to new scenarios without re-training. To mitigate this limitation, an exhaustive mapping of possible system responses under different operating conditions and values of model parameters is required. In practice, in high-dimensional calibration problems with systems operating under a large range of conditions, an exhaustive mapping is infeasible. In addition, such methods exhibit poor performance in scenarios involving noisy observations, limiting their implementation in practical applications.

Where a real-world system's behavior is not well represented by a physics-based model's structure, a popular framework for model calibration is the \textbf{probabilistic framework} proposed by \cite{Kennedy2001}. In this framework, both the physics-based model response and the model discrepancy are modelled with Gaussian Processes (GP). While GP is an elegant solution to emulate the response of a physics-based model and is well suited for uncertainty quantification, the GP representation can a) limit the class of functions that can be modelled, b) restrict the scalability to large datasets \cite{Rasmussen2006}, and also c) suffer from poor extrapolation ability. Additional computational issues arise from the use of Markov chain Monte Carlo to perform inference. Several recent developments have been proposed to mitigate these limitations, including the extension of the modelling capabilities of GP with Deep GP \cite{damianou13a} or considering variational inference \cite{Marmin2018VariationalCO}. Although the representation of complex physics-based models with Deep GP reduces the scalability limitations of classical GPs, for large-scale calibration problems, the scalability and computational time at run time of Deep GP-based methods for real-time model calibration in real-world scenarios is still limited \cite{Marmin2018VariationalCO}.

Because of the issues mentioned above, the dynamic, real-time, robust, and accurate inference of physics-based model parameters of complex engineered systems remains challenging. However, recent developments in \textbf{model-free reinforcement learning (RL)} have fostered a great deal of progress in addressing related control problems \cite{Zhang2020}. In fact, RL has proven to be effective in finding optimal control policies for non-linear stochastic systems when the dynamics are either unknown or affected by severe uncertainty \cite{bucsoniu2018reinforcement}, including complicated robotic locomotion and manipulation \cite{kumar2016learning, xie2019iterative, hwangbo2019learning}. The policies learnt via RL have the ability to adapt to new scenarios and scale well to large-scale problems at run time. In fact, the decision-making of reinforcement learning takes place through a neural network without any further optimization, which overcomes the inference speed problem at deployment time. Therefore, model-free RL \cite{sutton1992reinforcement} is a compelling alternative to traditional inference methods for physics-based model calibration.

\begin{figure*}[h]
\centering
\includegraphics[width=12.5cm]{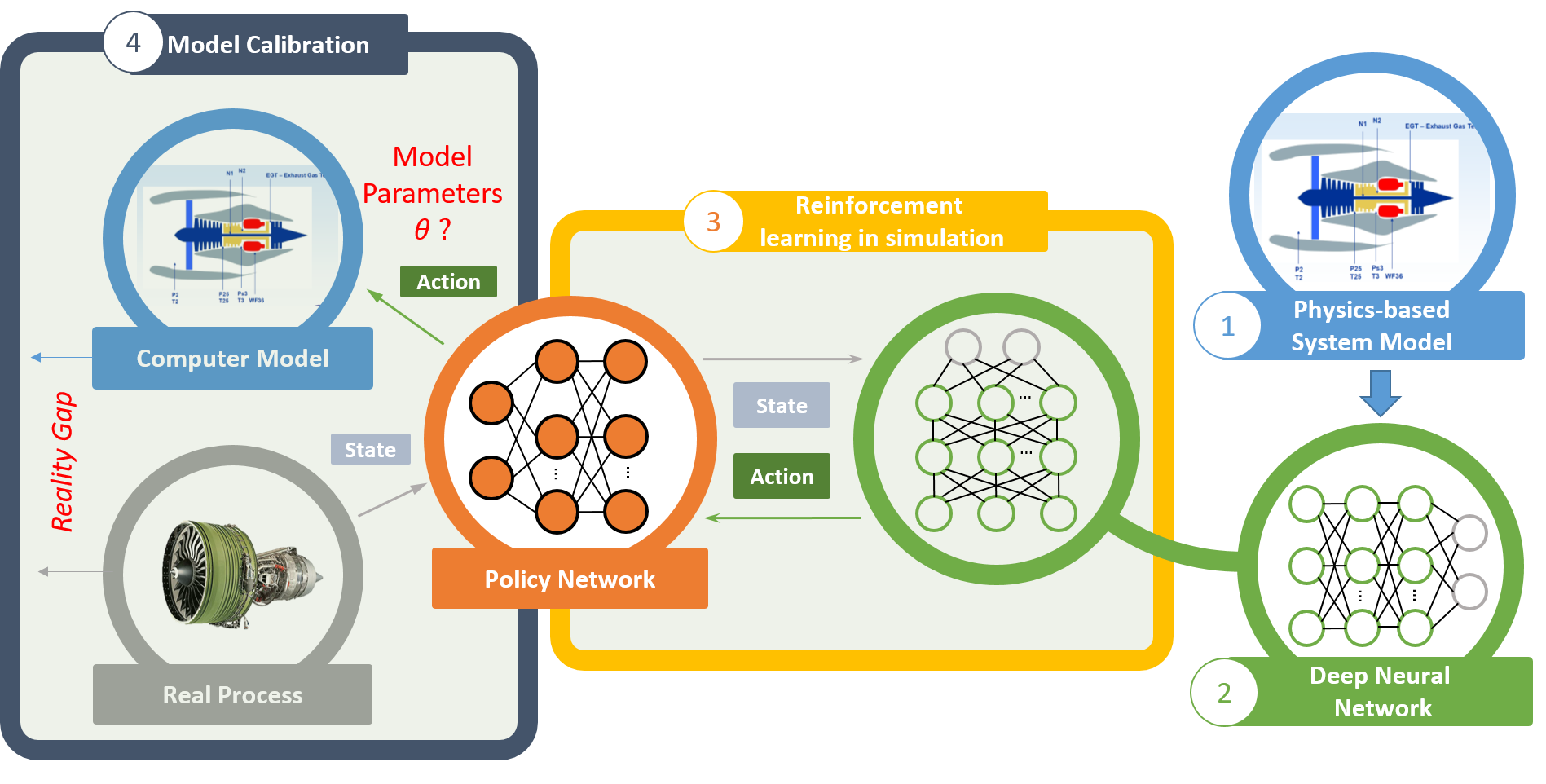}
\caption{\textbf{Creating a calibration policy:}  Step 1, we identify the parameters of the physics-based model we intend to calibrate. Step 2 (optional), we create a deep neural network (DNN) that models the complex system dynamics. Step 3, we train a control policy using the the physics-based model or the DNN model. \textbf{Implementation stage:} Step 4, we deploy the trained policy for real-time model calibration.}
\label{fig:framework}
\end{figure*}

One can realize the potential of utilizing model-free reinforcement learning for the inference of physics-based model parameters if one leverages the strong connection between inference in probabilistic models and reinforcement learning \cite{levine2018reinforcement}. In fact, as highlighted in \cite{levine2018reinforcement}, the connection between probabilistic inference and optimal control has been covered in the literature under different names: a) the Kalman duality \cite{Todorov2008}, b)  Kullback–Leibler (KL) divergence control \cite{Kappen2009}, c) stochastic optimal control \cite{Toussaint2009}, and d) maximum entropy reinforcement learning \cite{Ziebart2010}.

In this work, we propose a novel formulation of the calibration problem as a \textbf{tracking problem} that is modeled by a Markov decision process. Based on this formulation, we apply maximum entropy deep reinforcement learning to train an agent that controls the physics-based model parameters to keep the model response matching the observations. In order to achieve greater robustness to observation uncertainty and model inadequacy, we propose a novel constrained Lyapunov-based actor-critic (CLAC) algorithm. The proposed CLAC algorithm adds constraints on the stability of the policy network and is an extension of the Lyapunov-based actor-critic (LAC) algorithm.

Without any knowledge of the physics-based model or simulator, the agent is able to exploit the full dynamics of the model and produce robust control (i.e., calibration) logic. Therefore, the proposed framework overcomes the difficulties of traditional optimal control methods and data-driven approaches. It provides: a) accurate real-time dynamical calibration, b) a policy that can adapt to new scenarios without having been specifically trained on them, c) scalability to large datasets and high-dimensional spaces, and d) robustness to observation and model uncertainty.

The proposed framework is summarized in Figure \ref{fig:framework}. In the first step we identify the parameters of the physics-based model that are subjected to inference. In a second step, we use a physics-based model or, alternatively, a deep neural network (DNN) model that emulates the expected system response for measured properties (i.e., observations). In the third step, we use the DNN model to train the calibration policy network via RL. At deployment time, the trained calibration policy is directly deployed to obtain the physics-based model parameters at run time (step 4). The resulting calibration policy is computationally efficient at run time. Most importantly, the calibration policy is robust to uncertainty in the observations and the physics-based model. The proposed methodology is demonstrated and evaluated on a model-based diagnostics test case utilizing two different physics-based models of a turbofan engine: the Advanced Geared Turbofan 30,000 (AGTF30) and Commercial Modular Aero-Propulsion System Simulation (C-MAPSS) from NASA.

The contribution of this paper is two-fold: 1) We propose a solution to the problem of \textbf{real-time dynamic calibration of physics-based models}. In particular, we present a very general reinforcement-based model calibration framework that enables real-time inference of system model parameters without any online optimization and could be easily implemented on any system model. 2) From the methodological perspective, we propose the \textbf{constrained Lyapunov-based actor-critic (CLAC) algorithm}, which provides more action stability, especially on parameter tracking problems, compared to the state-of-the-art LAC reinforcement learning algorithm. This makes the proposed approach robust to noise and high variability.

\section{Background}
In this section, we briefly review the basic concepts and notations related to physics-based model calibration and reinforcement learning as they are the building blocks of the framework and method proposed in this work. In addition, we briefly introduce two traditional calibration approaches (unscented Kalman filters and end-to-end mappings with deep neural networks) to which we compare the performance of our proposed methodology.

\subsection{Calibration of physics-based models} 
The problem of calibration of physics-based models corresponds in its general form to the problem of modelling a physical process as approximated by a physics-based model. Observations of the real system response are given in the form of sensor readings $x_{t} \in R^{p}$ taken at variable inputs $w_{t} \in R^{s}$ representing, for instance, the operating conditions at time $t$. The physics-based model $F(w_{t},\theta_{t})$ provides approximations of the real process $\hat{x}$ at input condition $w_t$ given some values of the calibration inputs $\theta_{t} \in R^{d}$. Model calibration aims to infer the (unknown) value of $\theta_t$ that makes the model predictions follow the observations, i.e., $\hat{x} \sim x$. Following the formulation in \cite{Marmin2018VariationalCO}, the calibration problem can be generalized as:
\begin{equation}
 x_{t} = g(F(w_{t},\theta_{t}),w_{t})
\label{eq:cal}
\end{equation}
In this formulation, the observations are the result of an unknown stochastic warping mapping $g$ over the system model and the inputs $w_t$. It is worth pointing out that the original formulation in \cite{Kennedy2001} is obtained when $g$ applies the identity to $F(w_t,\theta_t)$ and the mismatch between the physics-based model and the reality (i.e., model discrepancy $\delta(w_t)$) is modelled by the warping function over the input variables $w_t$:
\begin{equation}
 x_{t} = F(w_{t},\theta) + \delta(w_{t})
\label{eq:cal2}
\end{equation}

\subsection{Reinforcement Learning}
Reinforcement learning is a sub-field of machine learning that focuses on how an agent interacts with the environment to achieve a specific goal. The environments are typically stated in the form of a Markov decision process (MDP), which provides a mathematical description of decision-making processes. Under the right problem formulation, MDPs can be useful for solving optimization and inference problems, such as the one described above for physics-based model calibration, via reinforcement learning. The details of the MDP formulation of physics-based model calibration will be discussed in Sec \ref{sec:main result}.

In conventional reinforcement learning, an agent is trained to interact with the environment and seek rewards on the basis of its actions. The agent receives a successor state $s_{t+1}$ from the environment as feedback in response to a decision (i.e., action $a_t$) taken at time-step $t$. The goal is to find a policy $\pi$ that maximizes the discounted cumulative reward $J(\pi)$ \cite{sutton1998introduction}, which is given by the following expression:
\begin{equation}
    J(\pi) = \mathbb{E}_{\tau \sim \rho_{\pi}}{\sum_{t=0}^{\infty} \gamma^t r(s_t, a_t)}
\label{eq:j}
\end{equation}
where $\gamma \in [0,1)$ is the discount factor.

\textbf{Maximum entropy RL.} The maximum entropy reinforcement learning framework considers a more general objective, aiming to learn a stochastic policy which jointly maximises the expected discounted cumulative reward and its expected entropy $\mathcal{H}(\pi(\cdot|s_t))$ \cite{ziebart2010modeling}:
\begin{equation}
    J(\pi) = \mathbb{E}_{\tau \sim \rho_{\pi}}{\sum_{t=0}^{\infty} [r(s_t, a_t)}+\beta\mathcal{H}(\pi(\cdot|s_t))], 
\label{eq:merl}
\end{equation}
where $\beta$ is the temperature parameter that controls the stochasticity of the optimal policy over the reward. Therefore, the resulting stochastic policies balance the exploration-exploitation trade-off and add robustness to the policy. Soft Actor-Critic (SAC) \cite{haarnoja2018soft} is one of the state-of-the-art off-policy reinforcement learning algorithms based on the maximum entropy reinforcement learning framework.

\textbf{Stability guaranteed RL.} The maximum entropy reinforcement learning framework can also include a closed-loop stability guarantee of the system dynamics. Such a stability guarantee is particularly relevant when dealing with control problems in real-world applications. Recently, the Lyapunov-based actor-critic (LAC) method \cite{tian2019model}, implementing a stability guarantee, showed state-of-the-art performance on tracking tasks. From a control-theoretic perspective, the task of tracking can be addressed ensuring that the closed-loop system is asymptotically stable. In other words, starting from an initial point, the trajectories of states always converge to a single point or reference trajectory. Therefore, in \cite{tian2019model}, a stability-guaranteed reinforcement learning framework is proposed under the following definition of stability:

\textit{Stability Definition.} Suppose $c_\pi(\cdot)$ is the cost function, $c_\pi : \mathcal{S} 
\to \mathbb{R}_+$. The system is said to be mean square stable (MSS) if 
$\lim_{t\rightarrow \infty }\mathbb{E}_{s_{t}} c_\pi(s_{t})=0$ holds for any initial condition $s_{0}$. 

Under this definition, the stability objective is given by Equation ~\ref{Theorem}. The stability objective defines an energy decreasing condition that drives the trajectory asymptotically to the null space of the cost function, producing predictable behaviour of the agent. Here, we use the Lyapunov function to denote the system's energy, so that the state goes in the direction of decreasing the value of the Lyapunov function and eventually converges to the origin or a sub-level set of the Lyapunov function. 
\begin{equation}
\mathbb{E}_{s\sim \tau }(\mathbb{E}_{s^{\prime }\sim P_{\pi }}L(s^{\prime
})-L(s))\leq -\alpha_{3} \mathbb{E}_{s\sim \tau }c_\pi\left( s\right)
\label{Theorem}
\end{equation}
where the $\alpha_3$ term controls the energy decreasing speed. 

\subsection{State-Update Method: Unscented Kalman Filter} 
Estimation of the physics-based model parameters from a transient data stream can be addressed with a traditional state-space formulation. In this solution strategy, the state vector comprises the model parameters and is modelled as a random walk. The measurement equation depends on the states and the input signals at the present time step $t$, which is available from the system model $F$. Under this formulation, a UKF can be applied to a non-linear discrete time system of the form:
\begin{align} \label{eq:ukf}
      \theta_{t} &= \theta_{t-1} + \xi_{t}  \\
       \hat{x}_{t} &= F(w_{t}, \theta_{t}) + \epsilon
\end{align}
where $\xi \sim N(0,Q)$ is a Gaussian noise with covariance $Q$ and $\epsilon \sim N(0,R)$ is a Gaussian noise with covariance $R$.

\subsection{End-to-End Learning}
An alternative approach to the calibration problem is to define a supervised learning set-up aimed at discovering a direct mapping from the condition monitoring data $[w,x]$ to the target $\theta$. Different machine learning can be applied for this task. This approach is valid under the assumption that the training dataset is representative of the testing dataset. In this case, the supervised models can generalize well on the test set. However, the extrapolation capabilities of such approaches are limited, which can be a significant limitation for real applications in evolving environments. 

The end-to-end learning strategy requires one to train a neural network in the inverse relation to the measurement equation of a state-update method:
\begin{align} \label{eq:e2e}
      \theta_{t} &= G(w_{t}, x_{t}, x_{t-1})
\end{align}
Since it is a supervised learning setup, this approach calls for an initial training set with the ground truth for the calibration parameters that are used as labels. This would require solving the inverse problem by other methods, using the results of other calibration methods as labels for the learning approaches or using synthetically generated labels. These are a crucial limitations of the end-to-end learning approaches.  

\section{Proposed Framework - Calibration Policy}\label{sec:main result}

\subsection{Model calibration defined as tracking problem}\label{sec:problem formulation}
In this work, we formulate the real-time model calibration problem as a tracking problem, which is modelled by an MDP, and use reinforcement learning to find the optimal tracking policy. The rationale behind this solution strategy is that learning to track observations of a real system response ($x_{t}$) by changing the model parameters ($\theta_{t}$) results in a control policy that makes the physics-based model yield a sound approximation of the physical process ($\hat{x}_{t}$), i.e., reducing the reality gap. Consequently, the tracking policy also serves as a calibration policy. It is worth noticing that this formulation of the calibration problem involves a system identification problem by tracking \cite{Ljung1990}.

Under a tracking solution strategy, the MDP describing the problem is given as the tuple ($S,A,r,P,\rho$), where the set of states($S$) comprises the current model output $\hat{x}_{t}$, the target value of the system response (observations of the real system) $x_{t+1}$, and the operating conditions $w_{t+1}$, i.e., $s_t=[\hat{x}_{t}, x_{t+1}, w_{t+1}]$. The set of actions ($A$) defines the model parameters that need to be calibrated, i.e., $a_{t}=\theta_{t}$. The reward/cost function $r(s,a)$ evaluates how good the tracking is. The state transition probability function ($P (s'|s,a)$) corresponds to the dynamics of the system that can be modelled by a physics-based model or surrogate model. 

\begin{figure}[h]
\centering
\includegraphics[width=10.5cm]{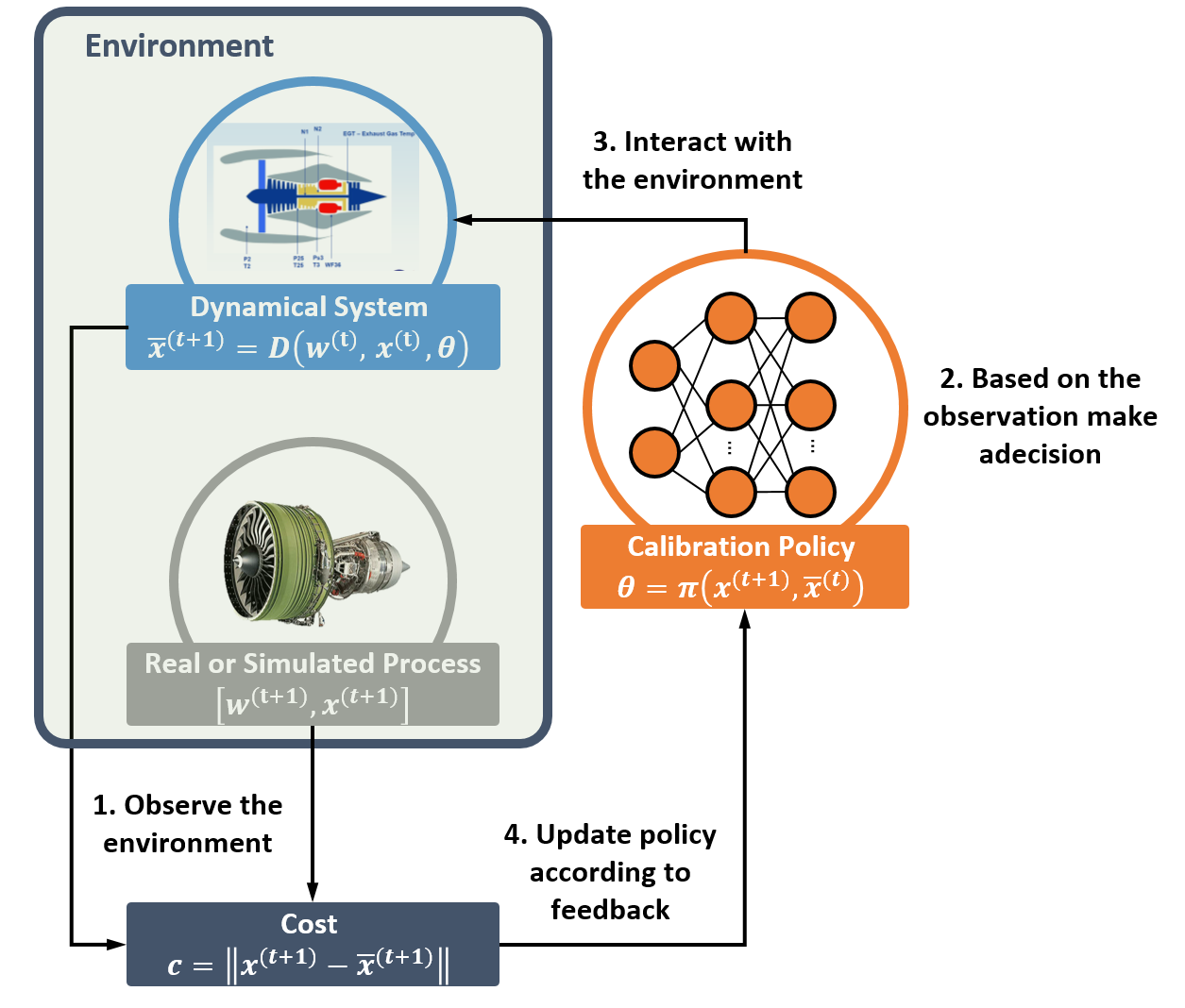}
\caption{Overview of the reinforcement learning loop. \textbf{Training stage:} 1) The agent observes the current state, which is described as $s_t=[\hat{x}_{t}, x_{t+1}, w_{t+1}]$ . 2) Then the agent makes a decision about the action $a_t=[\theta_{t+1}]$ according to the observation. 3) Following the interaction, the environment undergoes a change and the observation becomes  $s_{t+1}=[\hat{x}_{(t+1)}, x_{t+2}, w_{t+2}]$. 4) The agent receives a reward or loss according to his last action and updates his policy with this feedback. \textbf{Implementation stage:} 1)  Make the best decision based on the observation}.
\label{fig:Loop}
\end{figure}

In order to speed up the learning process of the RL algorithm, a discrete time counterpart of the physics-based model $F$ is used. The resulting dynamical system $D$ or simulator is modelled by a deep neural network that approximates the dynamic transition equation describing how the expected system response changes given the current observations $x_{t}$, the control variables $w_{t+1}$, and model parameters $\theta_{t+1}$, resulting in:
\begin{align} \label{eq:discrete_system}
 \hat{x}_{t+1}=D(w_{t+1}, \hat{x}_{t}, \theta_{t+1}) 
\end{align}
For the tracking problem there is, therefore, a desired state that we would like the system to be in at each time step, i.e., $x_{t+1}$. The task of the agent is to find a control policy $\theta_{t+1} = \pi(\hat{x}_{t}, x_{t+1}, w_{t+1})$ that minimizes the cost based on a specific distance metric representing the reality gap of the physics-based model. In particular, given the dynamical system above and a target system trajectory (i.e, observations), we train the control policy to keep the simulator state matching the real system state by maximizing the cumulative reward as given in Equation \ref{eq:j}. The complete reinforcement learning loop is shown in Figure \ref{fig:Loop}.

\subsection{Learning Algorithm}

In this work, we adopt Lyapunov-based actor-critic (LAC) \cite{tian2019model} as the learning algorithm, which is based on the soft actor-critic (SAC) \cite{haarnoja2018soft} algorithm and also incorporates a stability guarantee objective. The stability guarantee objective enables a control policy that stabilizes the system in the case of interference by unseen disturbances or uncertainties in the system dynamics. Most importantly, the LAC algorithm yields the best performance on tracking problems \cite{tian2019model}.

Based on the maximum entropy actor-critic framework, LAC uses the Lyapunov function $L_c$ as the critic in the policy gradient formulation. The objective function of $J(L_c)$ is given as follows:
\begin{equation}
    J(L_c) = \mathbb{E}_{(s,a)\sim \mathcal{D}}\left[\frac{1}{2}(L_c(s,a)-L_c^{target}(s,a))^2\right]
\end{equation}
\begin{equation}
    L_c^{\text{target}}(s,a)=c+\gamma \argmax_a L_c(s',a)
\end{equation}
where $L_c^{target}$ is the approximation target for $L_c$ as typically used in RL methods \cite{mnih2015human,lillicrap2015continuous}. $L_c^{target}$ has the same structure as $L_c$, but the parameter is updated through exponentially moving average of weights of $L_c$ controlled by a hyperparameter $\tau$. 

The objective function for the policy network is given by:
\begin{equation}
\begin{aligned}
J(\pi) = &\mathbb{E}_{ \mathcal{D}}[ \beta [\log(\pi_\theta(f_\theta(\epsilon,s)|s))]+ \\   
& \lambda(L_c((s',f_\theta(\epsilon,s')) - L_c(s,a)+\alpha_3c)]
\label{LAC}
\end{aligned}
\end{equation}
where $\pi_\theta$ is parameterized by a neural network $f_\theta$, and $\epsilon$ is an input vector consisted of Gaussian noise. The $\mathcal{D}\doteq\{(s,a,s',c)\}$ is the replay buffer for storage of the MDP tuples. In the above objective, $\beta$ and $\gamma$ are positive Lagrange multipliers which control the relative importance of policy entropy versus the stability guarantee. As in \cite{haarnoja2018soft2}, the entropy of policy is expected to remain above the target entropy $\mathcal{H}_t$. The value of $\beta$ is adjusted through gradient method, thereby maximizing the objective:
\begin{gather}
J(\beta) = \beta\mathbb{E}_{(s,a)\sim\mathcal{D}}  [\log(\pi_\theta( a|s))+\mathcal{H}_t]\label{eq:temperature update}
\end{gather}
and the $\lambda$ is adjusted by the gradient method, thus maximizing the objective:
\begin{gather}
J(\lambda) = \lambda(L_c((s',f_\theta(\epsilon,s'))-L_c(s,a)+\alpha_3c)
\label{eq:gamma update}
\end{gather}
Under conditions of high sensor noise and simulator bias resulting from an incomplete representation of the system model (i.e., irreducible reality gap), the policy network can exhibit large variance. Such a situation is undesirable in many real-world applications where it is important to obtain a stable or smooth action over time. Therefore, in order to stabilize the action, we introduce the constrained Lyapunov-based actor critic (CLAC) algorithm, a modification of the LAC, which significantly improves the action stability under model uncertainty and sensor noise. In CLAC, the objective function has an additional term that aims to obtain a policy network that has similar optimal action when given a similar or near state ($s_{near}$) and is given by:

\begin{equation}
\begin{aligned}
J(\pi) = &\mathbb{E}_{ \mathcal{D}}[ \beta [\log(\pi_\theta(f_\theta(\epsilon,s)|s))]+ \\
&\lambda(L_c((s',f_\theta(\epsilon,s'))-L_c(s,a)+\alpha_3c) + \\
&\alpha ||\pi_{\theta}^*(s)-\pi_{\theta}^*(s_{near})||]
\label{CLAC}
\end{aligned}
\end{equation}

where $\alpha$ is a positive Lagrange multiplier, and $\pi_{\theta}^*(s)$ outputs the action with largest probability. In our case, we use the adjacent time space state $s_{t+1}$ or $s_{t-1}$ to approximate $s_{near}$.

The entire procedure for training the proposed constrained Lyapunov actor-critic is provided in Algorithm \ref{algo:CLAC} and all the hyper-parameter settings may be found in the Appendix.

\begin{algorithm}[tb]
   \caption{ Constrained Lyapunov-based Actor-Critic (CLAC)}
   \label{algo:CLAC}
\begin{algorithmic}
    \STATE Input hyperparameters, learning rates $\alpha_{\phi_{L}}$, $\alpha_\theta$
   \STATE Randomly initialize a Lyapunov network $L(s, a)$ and policy network $\pi(a|s)$ with parameters $\phi_{L}$,  $\theta$ and the Lagrange multipliers $\beta$, $\lambda$
    \STATE Initialize the parameters of target networks with $\overline{\phi}_{L}\leftarrow\phi_{L}$,  $\overline{\theta}\leftarrow\theta$
   \FOR{each iteration}
   \STATE Sample $s_0$ according to $\rho$
   \FOR{each time step}
   \STATE Sample $a_t$ from $\pi(s)$ and step forward
   \STATE Observe $s_{t+1}$, $r_{t}$ and store $(s_t,a_t,r_t,s_{t+1})$ in $\mathcal{D}$
  
   \ENDFOR
   \FOR{each update step}
   \STATE Sample minibatches of transitions from $\mathcal{D}$ and update $L$, $\pi$ and Lagrange multipliers with gradients

    \STATE Update the target networks with soft replacement:
          \begin{align}
            \overline{\phi}_{L} &\leftarrow \tau \phi_{L} + (1 - \tau) \overline{\phi}_{L} \notag \\
            \overline{\theta} &\leftarrow \tau \theta +(1 - \tau) \overline{\theta} \notag
          \end{align}
   \ENDFOR
   \ENDFOR

\end{algorithmic}
\end{algorithm}

\section{Case Study: Diagnostics of safety-critical systems}

\subsection{Introduction to model-based diagnostics}
Model-based diagnostics aims to detect, isolate, and explain anomalies in the behaviour of a system by finding health-related model parameters that approximate the observed system response. Diagnostics of safety-critical systems, such as aircraft engines, is an active research area \cite{Li2002, Fentaye2019} with a long history going back to the original work of \cite{Urban1973}. Because of the potentially catastrophic impact of failures in such systems, it is important to provide accurate and robust inference of the health-related model parameters but also to perform this task in real-time to promptly raise the alarm and take mitigation actions with minimal delay. Yet current model-based diagnostics methods only offer a compromise between speed, robustness, accuracy, and scalability.

\subsection{Experiments}
The proposed framework and method are demonstrated and evaluated on two datasets generated with two different physics-based models focusing on the diagnostics of safety-critical systems represented by turbofan engines. Each dataset explores different aspects of real-world calibration problems. Dataset \#1 corresponds to a one-dimensional calibration problem ($d=1$) under a wide range of real (i.e., noisy) flight conditions from a small fleet of ten units ($N=10$). With 6.7M samples, Dataset \#2 is a large dataset that explores complex failure modes affecting four components of the system simultaneously ($d=4$). Therefore, Dataset \#2 explores a calibration problem under complex system responses. In contrast to Dataset \#1, it contains only data from one single unit and, consequently, has a more limited range of operating conditions. An overview of the two calibration problems is provided in Table \ref{tb:datasets}.

%
\begin{table}[ht]
  \caption[Table caption text]{Overview of Dataset \#1 and Dataset \#2. Size of the dataset ($m$), dimension of the observation vector ($n$), dimension of the calibration vector ($d$), number of units ($N$), fault type, number of fault types ($C$), and ranges of the scenario-descriptor variables, i.e., $W$: altitude ($\text{alt}$), flight Mach number ($\text{XM}$), and throttle-resolver angle ($\text{TRA}$)}
  \begin{center}
    \begin{tabular}{|c||c|c|}  \hline
       Parameter      & Dataset \#1     &  Dataset \#2     \\ \hline \hline   
       Model Name     & C-MAPSS         &  AGTF30          \\ \hline   
       $m$            & 0.5M            &  6.7M            \\ \hline   
       $n$            & 20              &  8               \\ \hline
       $d$            & 1               &  4               \\ \hline
       $N$            & 10              &  1               \\ \hline
       Fault Type     & Continuous      &  Discrete        \\ \hline
       $C$            & 10              &  1315            \\ \hline
       Alt [ft]       & 35.0k - 10.0k   &  29.0k - 25.7k   \\ \hline
       XM [-]         & 0.75 - 0.26     &  0.74 - 0.67     \\ \hline
       TRA [\%]       & 87.8 - 23.6     &  82.4 - 69.1     \\ \hline
    \end{tabular}
  \end{center}
\label{tb:datasets}
\end{table}

The performance of the proposed CLAC method is evaluated and compared to two alternative calibration models: a unscented Kalman filter (UKF) and a supervised end-to-end mapping with deep learning algorithm (E2E). The evaluation also covers variants of Dataset \#1 designed to evaluate the robustness of the different methods to uncertainty in the observations and system model predictions.

\begin{figure*}[ht]
\centering
\includegraphics[width=0.24\columnwidth]{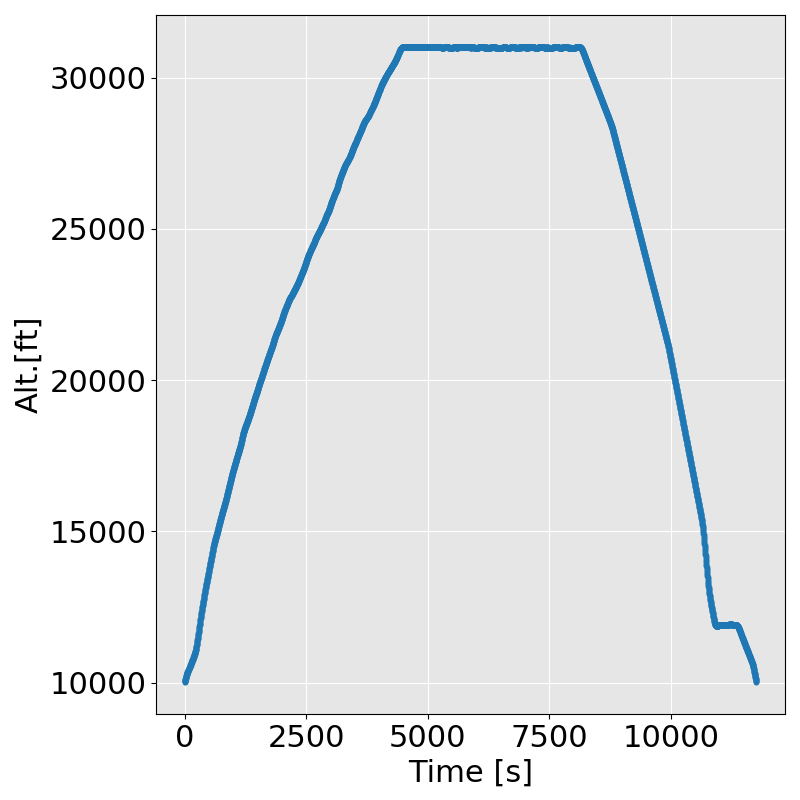}
\includegraphics[width=0.24\columnwidth]{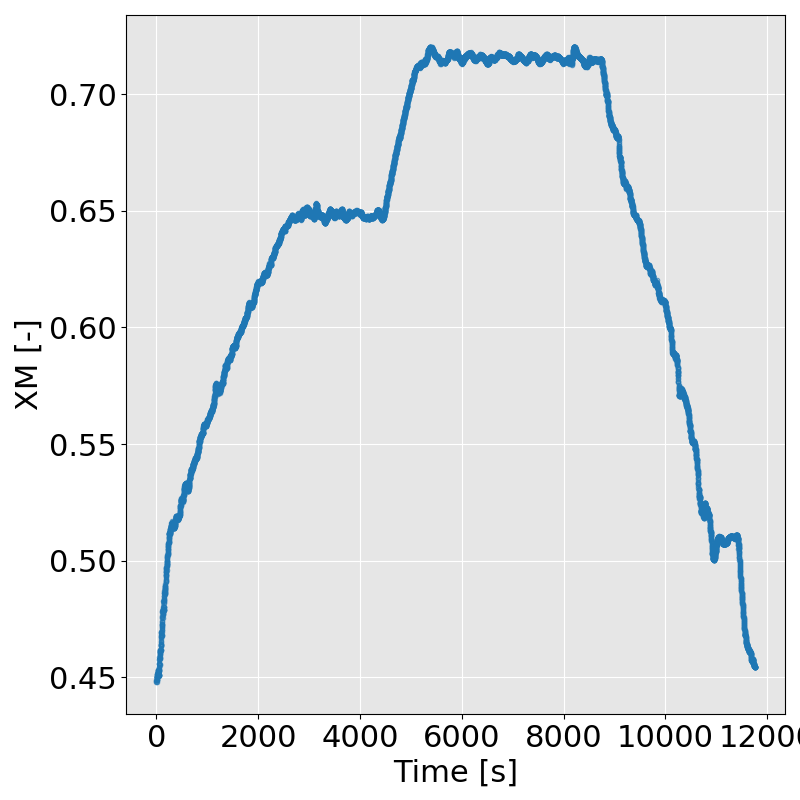}
\includegraphics[width=0.24\columnwidth]{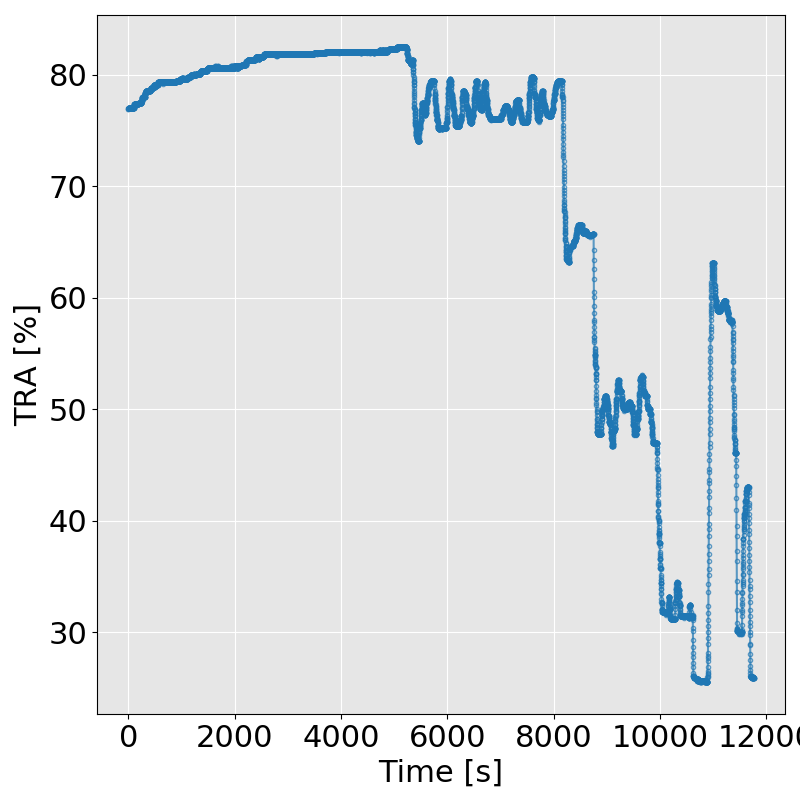}
\includegraphics[width=0.24\columnwidth]{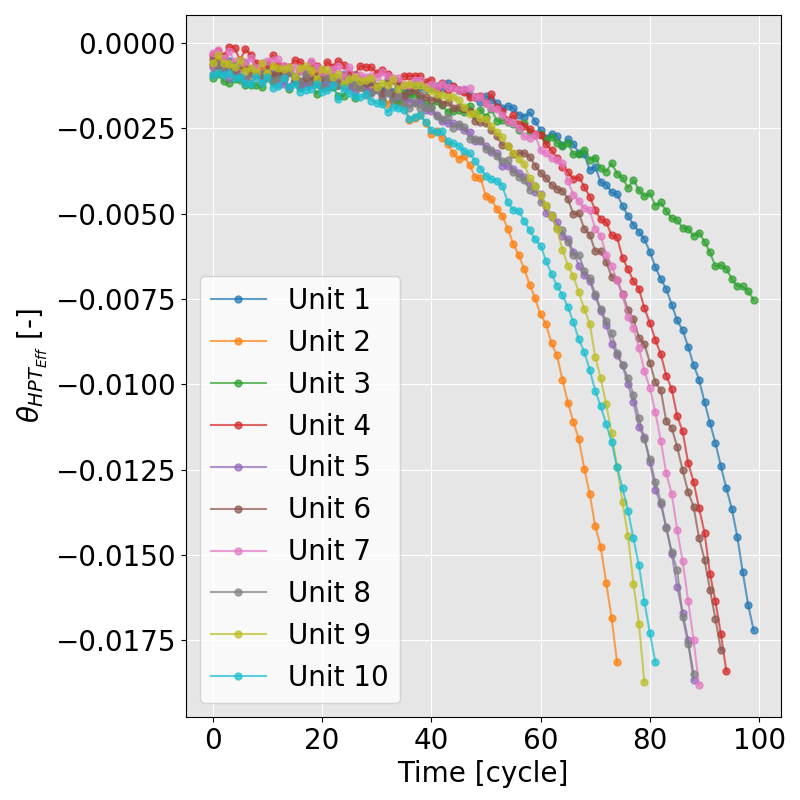}
\caption{\textbf{Left (1-3).} Typical flight profile given by the traces of altitude (Alt), flight Mach number (XM), and throttle-resolver angle (TRA) for Unit 2. The flight profile is restricted to $\text{alt}>10000$ ft and therefore covers climb, cruise, and descent flight conditions. \textbf{Right (4).} Ground truth degradation profiles of each unit of the fleet given by the trace of the model parameter HPT Eff. i.e., $\theta_{\text{HPT}_{\text{Eff}}}$ versus time in flight-cycles.}
\label{fig:CMAPSS}
\end{figure*}

\subsection{Dataset \#1: A Small Fleet of Turbofan Engines}
Dataset \#1 provides degradation trajectories of a small fleet comprising ten turbofan engines with unknown and different initial health conditions. The trajectories are given in the form of multivariate time-series of sensor readings (i.e., $[w, x]$). The dataset was generated with the Commercial Modular Aero-Propulsion System Simulation (C-MAPSS) dynamical model \cite{Frederick2007}. Real flight conditions ($w$), as recorded on board a commercial jet, were taken as input to the C-MAPSS model \cite{DASHlink}. Figure \ref{fig:CMAPSS} (left) shows a typical flight profile given by the scenario-descriptor variables ($w$): altitude ($\text{alt}$), flight Mach number ($\text{XM}$), and throttle-resolver angle ($\text{TRA}$) for ten units ($N=10$). All the units are affected by the same fault mode corresponding to the degradation of the high pressure turbine (HPT) efficiency. Figure \ref{fig:CMAPSS} (right) shows degradation profiles of each unit of the fleet given by the trace of the true HPT Eff. $\theta$. The degradation of the HPT evolves following a stochastic process with a linear \emph{normal degradation} followed by a steeper \emph{abnormal degradation}. The degradation rate of each component varies within the fleet. More details about the generation process can be found in \cite{AriasChao2020}.

As discussed above, generation on an supervised end-to-end deep learning model requires access to the ground truth labels i.e., $\theta$. Therefore, for training the E2E algorithm, we assumed that the labels $\theta$ are available for a subset of the units (Unit 1, 4, 7 \& 9) corresponding to low altitude and short flights. This experiment design generates a training dataset that is not fully representative of the possible system responses present in the test set where higher altitude and longer flights are present.

\subsection{Dataset \#2: A set of fault scenarios in turbofan engines}
Dataset \#2 provides simulated condition monitoring data (i.e., $[w,x_s]$) of an advanced gas turbine during three flight profiles and multiple fault scenarios. The dataset was synthetically generated with the AGTF30 (Advanced Geared Turbofan 30k lbf) dynamical model \cite{Chapman2017} taking as input real flight conditions as recorded on board a commercial jet \cite{DASHlink}. Concretely, three different flight trajectories with a duration of 5000 s are considered. The dataset consists of concatenated time series of sensor readings (i.e., $[w,x_s] \in R^n$) resulting from faulty engine conditions. The fault conditions are induced and simultaneously affect four health-related model parameters representing model modifiers of the high pressure turbine (HPT) and low pressure turbine (LPT) flow and efficiency. A total of 1315 different fault scenarios are generated by factorial design of a finite set of possible degradation intensities for each component. No additional noise was added to the model response since the flight conditions are already noisy.

\section{Results}
The aim of the proposed framework is to enable for the first time accurate, real-time, and robust model calibration for large-scale problems. Therefore, in this section, the performance of the proposed method is analysed based on six evaluation criteria: inference accuracy, computational cost, robustness to system model uncertainty, robustness to observation noise, scalability to large datasets, and tracking accuracy.

\begin{figure*}[h]
    \centering
    \includegraphics[width=0.32\columnwidth]{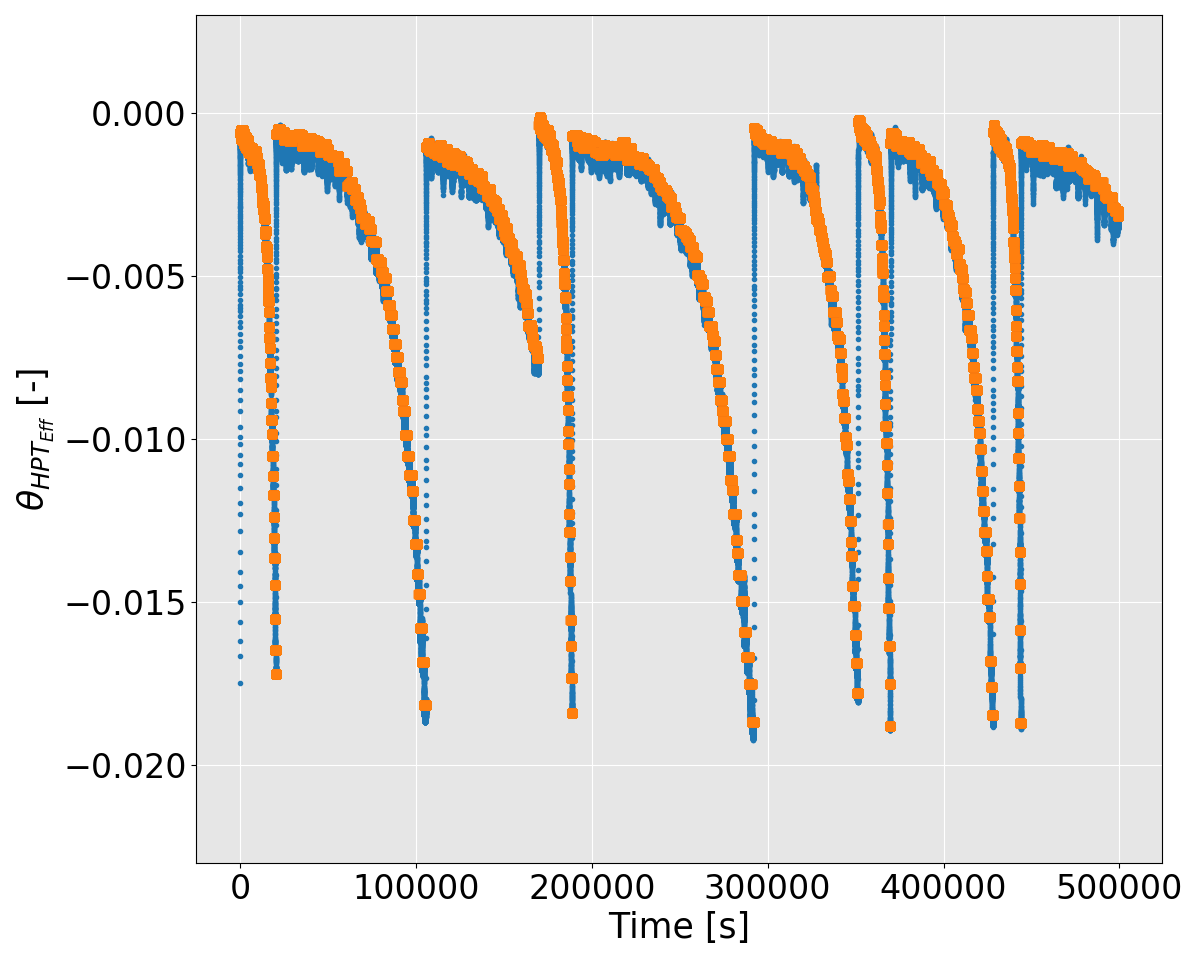}
    \includegraphics[width=0.32\columnwidth]{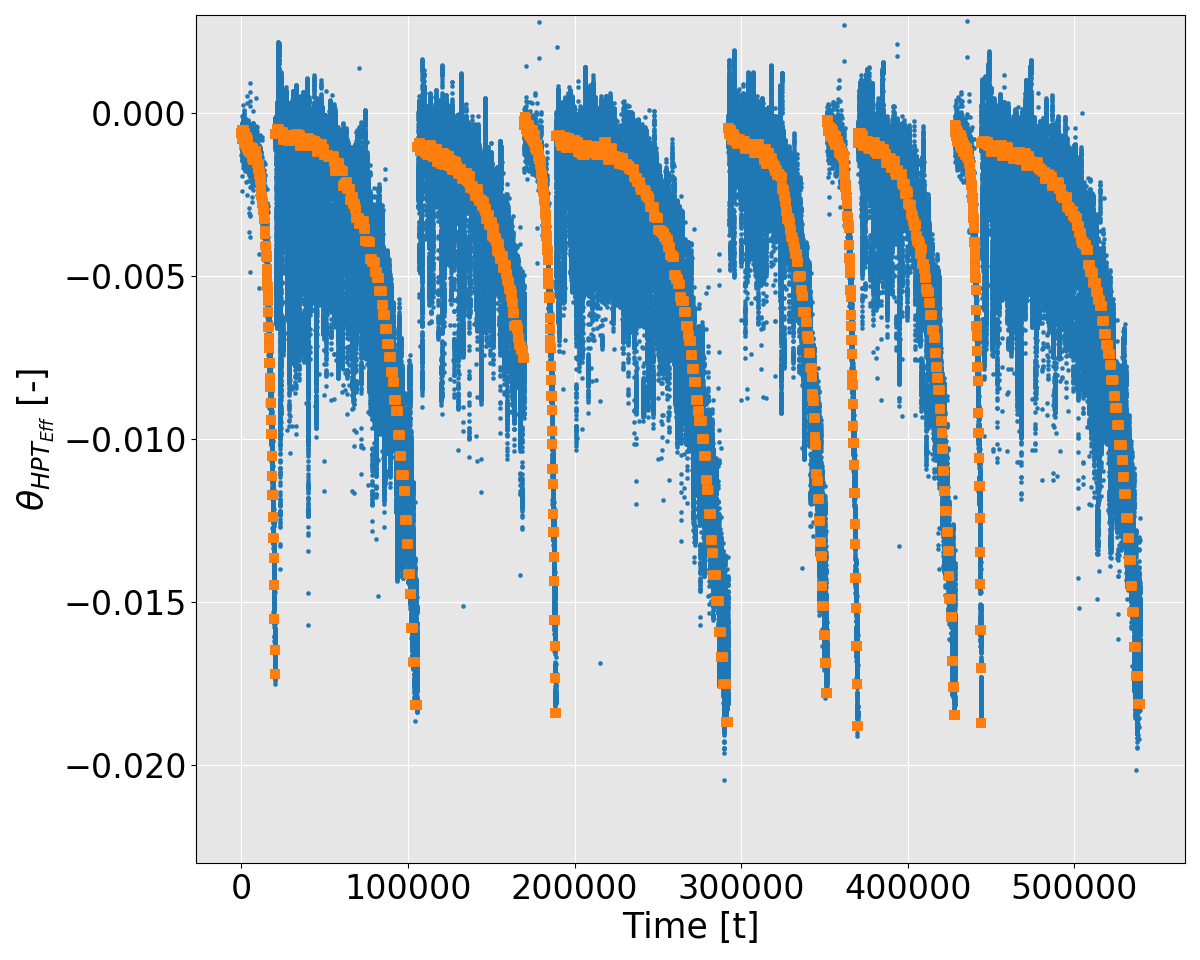}
    \includegraphics[width=0.32\columnwidth]{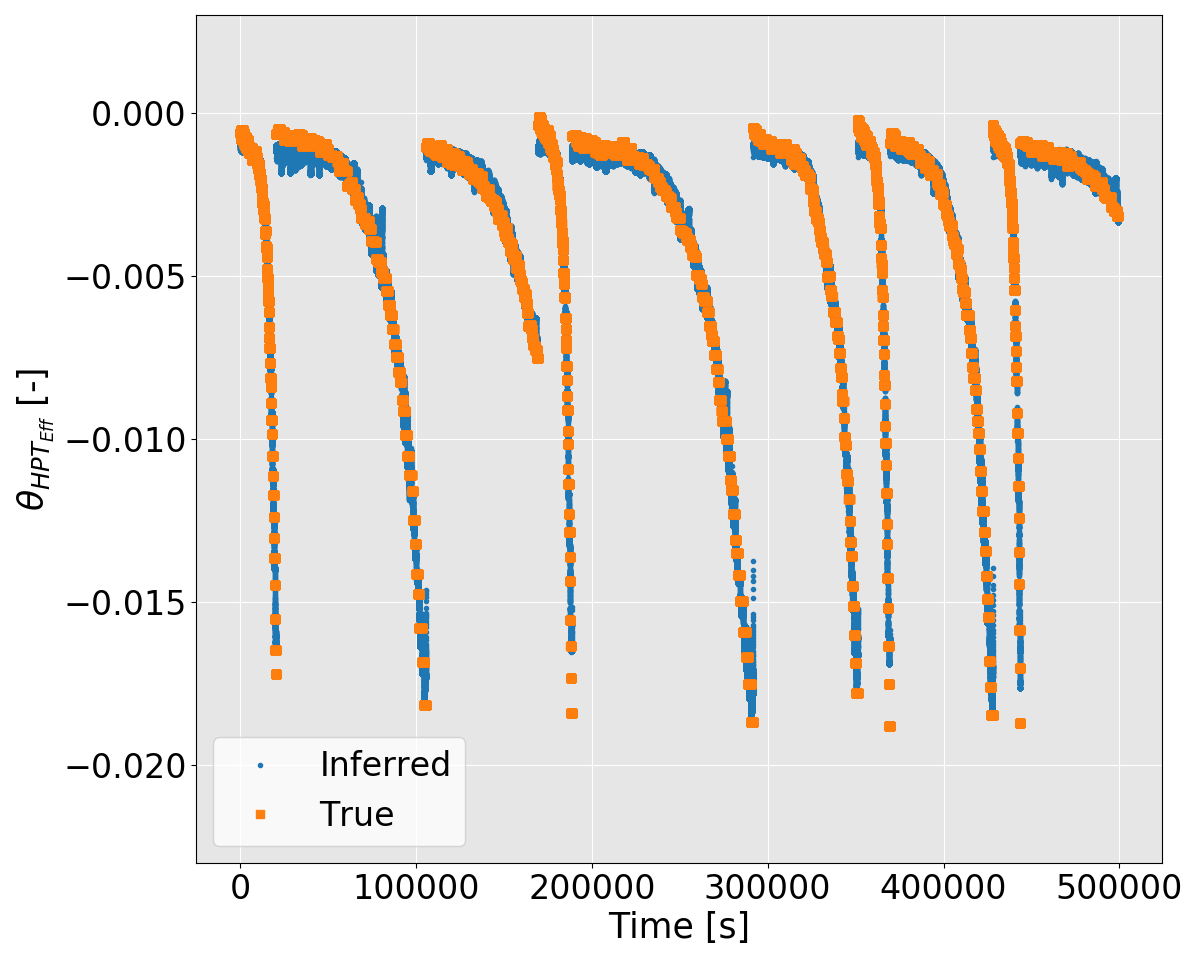}
    \caption{Inferred (blue dots) and ground truth (orange squares) traces of $\theta$ in Dataset \#1 with UKF (left), E2E (middle), and CLAC (right) approaches. The $\theta$ values for the ten units are stacked one after the other, generating a single time sequence. Each discontinuity corresponds to the beginning of a new unit. The UKF solution proves a good match to the ground truth, with low bias and variance. It is observed that only at the beginning of each unit do the UKF predictions show large bias. Estimations with the E2E model show large variance, in particular for units with long degradation profiles. The CLAC method shows itself to be a very good match to the ground truth.}
    \label{fig:Inference_Comp}
\end{figure*}

\textbf{Inference Accuracy.} The primary objective of model calibration is to infer the values of the model parameters $\theta$. From the application perspective of model-based diagnostics, this objective corresponds to inferring the true underlying degradation parameters. Therefore, we compare the estimated degradation parameters ($\hat{\theta}$) with the ground truth and report the inference accuracy in the form of the root mean square error (RMSE). Table \ref{tb:inference} shows the inference performance of the unscented Kalman filter (UKF), end-to-end mapping (E2E), and the proposed method (CLAC) in both datasets. With the lowest RMSE, the policy obtained with CLAC shows the best overall performance in both datasets. The improvement is particularly significant under complex fault modes (i.e., Dataset \#2). The E2E model yields the worst overall performance in Dataset \#1, which highlights the limitations of supervised learning in cases where the training dataset is not fully representative of the test conditions. Figure \ref{fig:Inference_Comp} shows the inferred unobserved model parameters $\hat{\theta}$ obtained with the three methods in Dataset \#1. It is worth mentioning that unlike the end-to-end mapping, which needs the ground truth degradation parameters for training, our framework does not need any prior knowledge about the degradation parameters. This makes the approach more flexible and more applicable to real scenarios. 

\begin{table}[ht]
\caption[Table tb:inference]{Overview of the inference performance given by the RMSE between the inferred model parameters and the ground truth with UKF, E2E, and CLAC approaches on complete test trajectories. Best performance is shown in \textbf{bold}. $^1$ The analysis with the E2E model was not performed in Dataset \#2 in light of the poor results with Dataset \#1}
  \begin{center}
    \begin{tabular}{|c||c|c|}
      \hline
       Method       & Dataset \#1         &  Dataset \#2       \\ \hline
       UKF          & 3.42e-04            &  3.51e-03          \\ \hline
       E2E          & 1.36e-03            &  $-^1$             \\ \hline   
       CLAC         & \textbf{3.30e-04}   &  \textbf{2.50e-03} \\ \hline   
    \end{tabular}
  \end{center}
\label{tb:inference}
\end{table}

\textbf{Computational Cost.} One crucial aspect of the proposed method is the ability to perform real-time calibration. Therefore, we evaluate the time required to perform inference of the model parameters at deployment. Table \ref{tb:time} reports the average times required to calibrate a single sample and the total training time with the three methods. In terms of deployment computational cost, the proposed method provides a speed up of $\times 1500$ compared to the UKF. Concretely, inference with the proposed CLAC method takes around 40 ms using a CPU thread. This deployment speed is comparable to the E2E model as both methods only require a forward pass over a deep neural network. By contrast, the UKF needs to perform $2 \times (2 \times d + 1)$ model evaluations, which for Dataset \#1  amounts to 6 s. The CLAC method requires several hours of training with an ordinary PC and therefore incurs all the computational cost in the training phase, which is typically not critical for practical applications. For real-time applications, the main limiting factor is the deployment time. Therefore, in terms of computational cost, the proposed method has a clear advantage over the current state-of-the-art approaches.

\begin{table}[ht]
\caption[Table tb:time]{Overview of the average time required for inference of a single sample and the total training time with UKF, E2E, and CLAC approaches in seconds [s] for Dataset \#1.}
  \begin{center}
    \begin{tabular}{|c||c|c|} \hline
       Method              &  Deployment Time [s] & Training Time [s] \\ \hline
       UKF                 & 6                    & \textbf{0}        \\ \hline
       E2E                 & 4.2e-04              & 200               \\ \hline
       CLAC                & \textbf{4.0e-04}     & 6200              \\ \hline
    \end{tabular}
  \end{center}
\label{tb:time}
\end{table}    

\textbf{Robustness to Environment Uncertainty.}
Robustness to model inaccuracy is an important aspect in model calibration. It is also a well known limitation of model-based methods such as UKF. To evaluate the sensitivity of different approaches to inaccuracies in the models, we apply a model bias to the output of the dynamic system model (i.e., $F(w,\theta) \times \alpha$) to emulate an inadequacy of the system model structure (i.e., inaccurate simulator). We also consider a case whereby Gaussian noise is added to the dynamic system model, i.e., $F(w,\theta) + \eta$ where $\eta \sim N(0,\alpha_{\eta})$. It is worth noticing that adding noise to the output of the simulator transforms the deterministic model into a stochastic system model. 

From the RL perspective, the presence of an inaccurate simulator is known as \textit{sim-to-real} transfer. In fact, \textit{sim-to-real} is always a critical problem in reinforcement learning since the agent is trained in a simulated environment which may be different from the real world. In our case, we use a surrogate DNN model to accelerate the training. Therefore, we have an unavoidable error between the DNN surrogate model $\hat{x}_{t+1}=D(w_{t+1},\hat{x}_t, \theta_{t+1})$ and the engine physics-based model. Then, even in the case where noise is not added, the agent needs to make decisions with noisy DNN model outputs $\hat{x}_{t}$ at every time step $t$.

\begin{figure}[h]
\centering
\includegraphics[width=0.45\columnwidth]{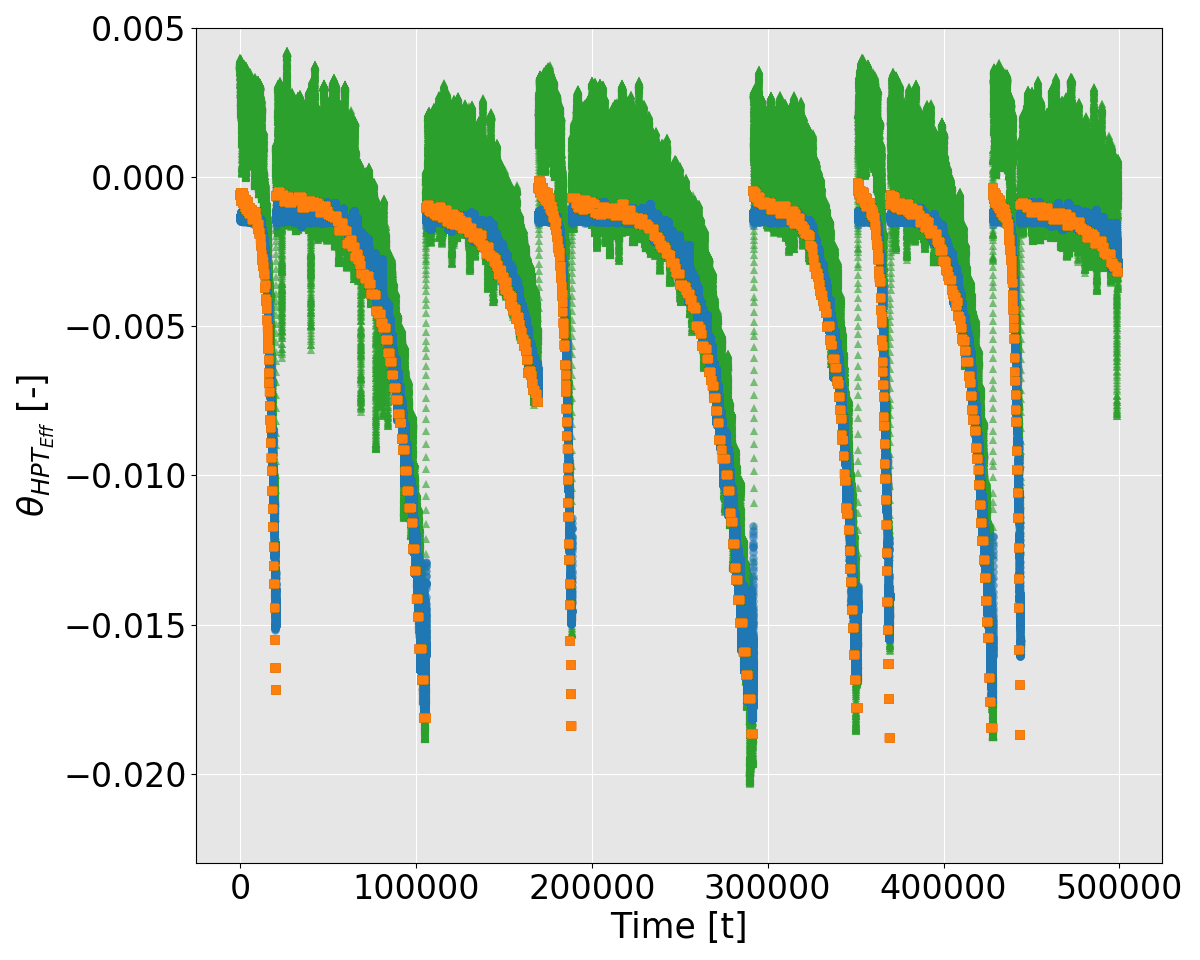}
\caption{Inferred and ground truth (orange squares) traces of $\theta$ in Dataset \#1 with 2\% model bias for UKF (green triangles) and CLAC (blue dots).}
\label{fig:Inf_Comp_bias}
\end{figure}

In order to test the trained policy under bias and noisy simulators, we tested two variants where we added a $2\%$ fixed bias (i.e., $\alpha=1.02$) and a 10\%  Gaussian noise. (i.e., $\alpha_{\eta}=0.1$) to the output of the DNN model. Table \ref{tb:robust_track} shows that the policy obtained with the CLAC model provides a very good inference even under quite large uncertainty, demonstrating better robustness than the UKF, which failed to optimize a stable inference. The superior inference performance of the CLAC model under $2\%$ fixed bias is visualised in Figure \ref{fig:Inf_Comp_bias}.

\begin{table}[h]
\caption[Table tb:time]{Overview of the inference performance (RMSE) under model bias ($F(w,\theta) \times \alpha$) and noise  ($F(w,\theta)+ \eta$) with UKF and CLAC approaches in Dataset \#1.}
  \begin{center}
    \begin{tabular}{|c||c|c|} \hline
      \multicolumn{3}{|c|}{ Model Bias: $F(w,\theta) \times \alpha$ } \\ \hline \hline
       Intensity             & UKF      &  CLAC               \\ \hline
       $\alpha =  1.02$      & 2.04e-3  &  \textbf{3.30e-04}  \\ \hline \hline 
    \multicolumn{3}{|c|}{ Model Noise: $F(w,\theta)+ \eta$}   \\ \hline \hline
       Intensity             & UKF      &  CLAC               \\ \hline
       $\eta \sim N(0,0.1)$  & \#       &  \textbf{4.22e-04}  \\ \hline 
    \end{tabular}
  \end{center}
\label{tb:robust_track}
\end{table}

\begin{figure*}[h]
\centering
\includegraphics[width=0.34\columnwidth]{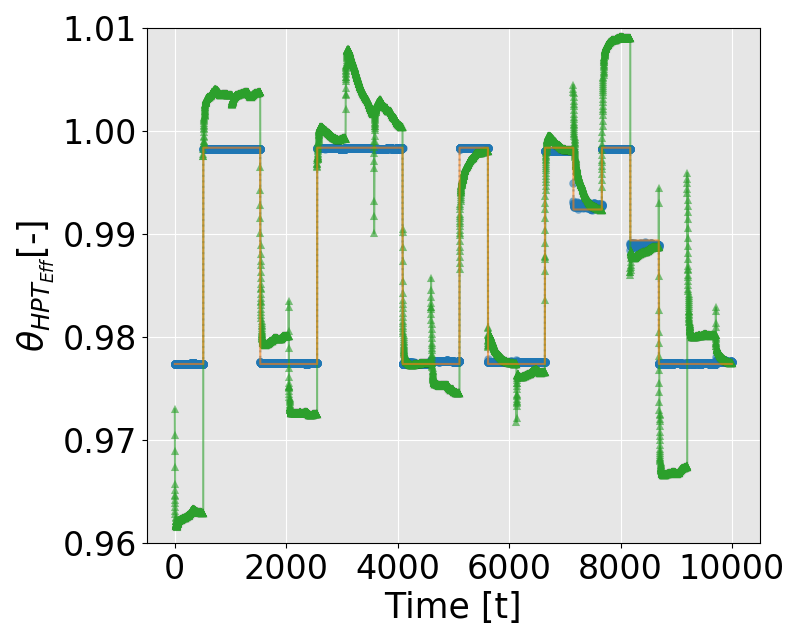}
\includegraphics[width=0.34\columnwidth]{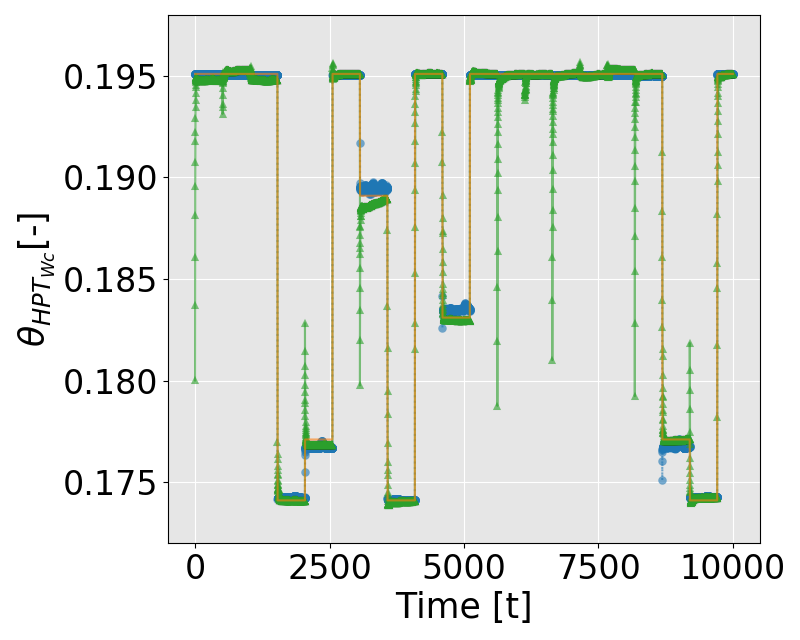}
\includegraphics[width=0.34\columnwidth]{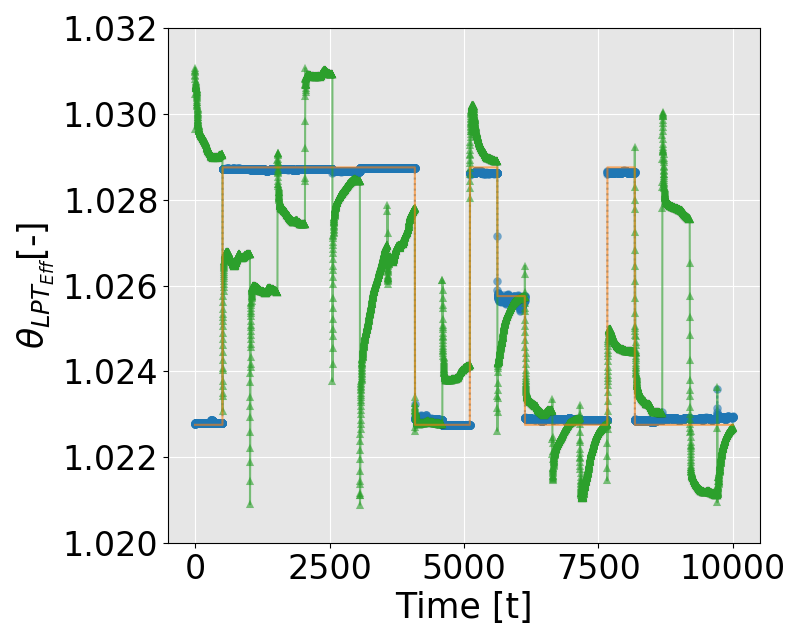}
\includegraphics[width=0.34\columnwidth]{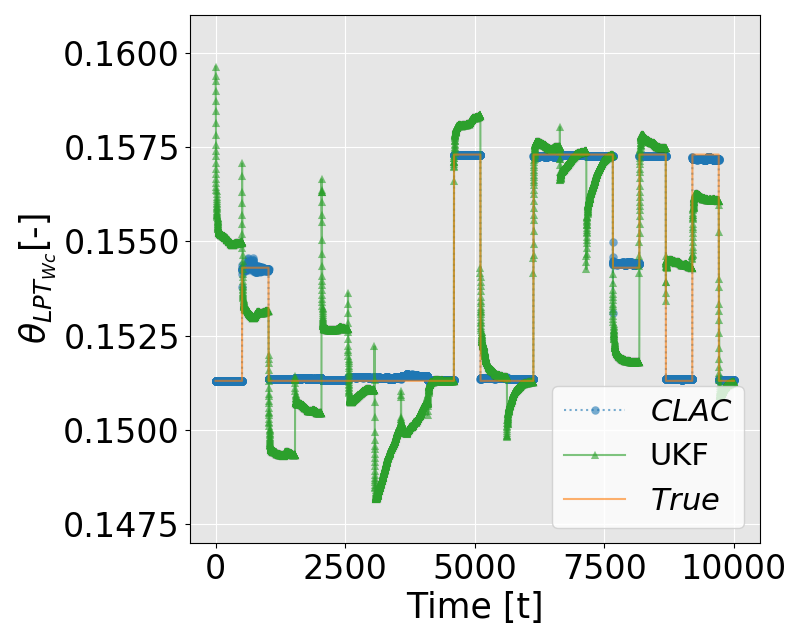}\\

\caption{Comparison to the ground truth in four $\theta$ (HPT and LPT flow (i.e., Wc) and efficiency (i.e., Eff.)) and different degradation parameter settings. The orange solid line is the ground truth degradation parameter value in different trajectories. The blue dotted line is the policy's action  with CLAC and the green solid line with triangles is the UKF prediction.}
\label{fig:Inference_AGT30}
\end{figure*}

\textbf{Scalability to Large Dataset and High Dimensional Model Calibration Parameters $\theta$.} When the dimensionality of the physics-based model parameters $\theta$ increases, the complexity of inference increases as well. Due to the non-linear correlation between the degradation parameters and also between the degradation parameters and observations, the solution of the calibration problem in high dimensional spaces can lead to \textit{confounding} solutions. In scenarios with noisy observations and systems with poor observability, the solution of inverse problems, such as UKF methods, might involve the spurious association of calibration factors that have similar system outputs. To test the scalability of our policies, we performed experiments on controlling 1, 2, and 4 degradation parameters in AGTF30 experiments (i.e. Dataset \#2). Figure \ref{fig:Inference_AGT30} shows the inferred and ground truth traces of a four-dimensional $\theta$ in Dataset \#2 with UKF (left) and CLAC (right) approaches. As in the previous plots, the $\theta$ values for 1315 fault intensities are stacked one after the other, thus generating a single time sequence. We can observe that the UKF solution does \textit{confound} or \textit{smear} the source of degradation. Moreover, as observed for Dataset \#1, at the beginning of each fault combination the predictions show large bias. Both of these issues are efficiently solved with the proposed CLAC method.

\begin{figure}[h]
    \centering
\includegraphics[width=0.34\columnwidth]{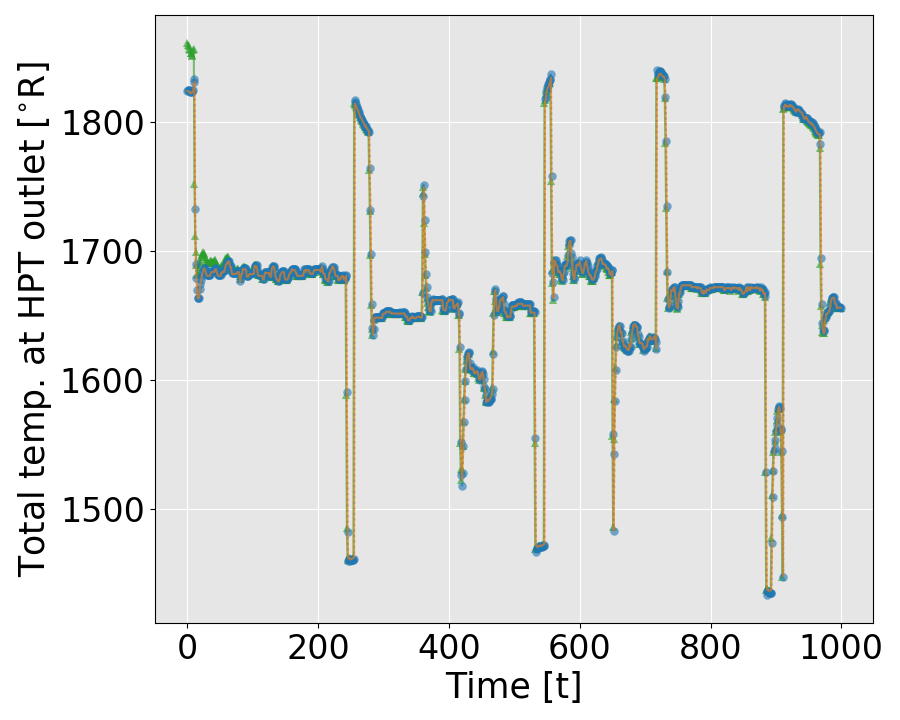}
\includegraphics[width=0.34\columnwidth]{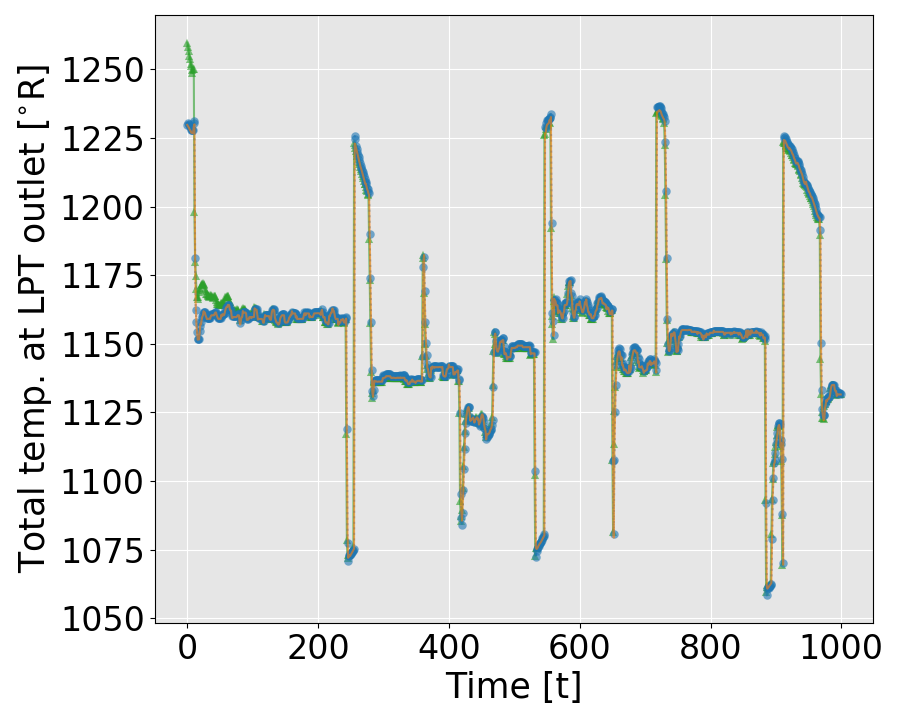}\\
\includegraphics[width=0.34\columnwidth]{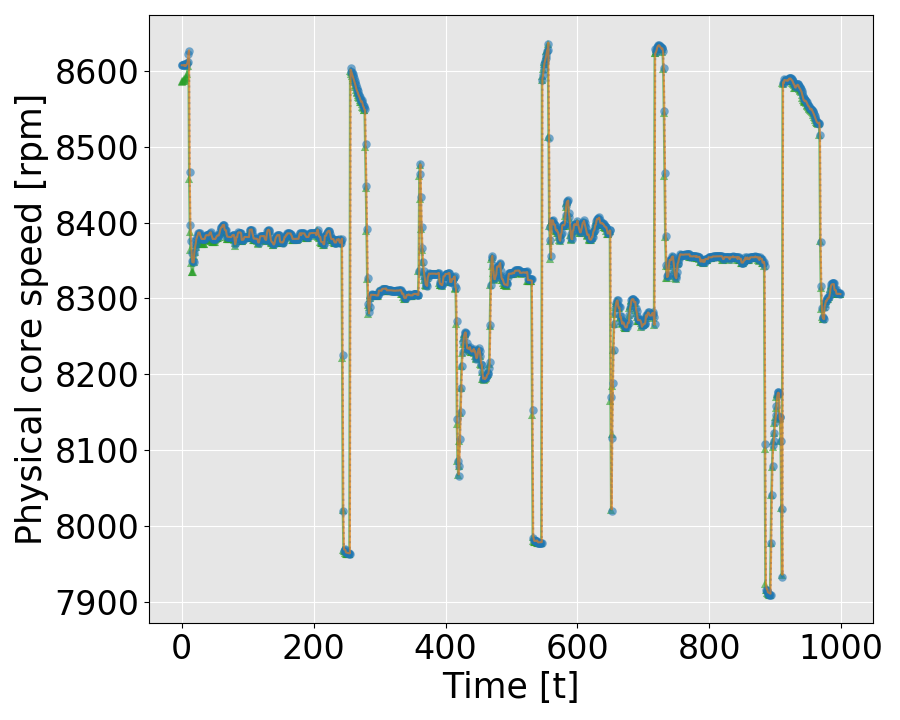}
\includegraphics[width=0.34\columnwidth]{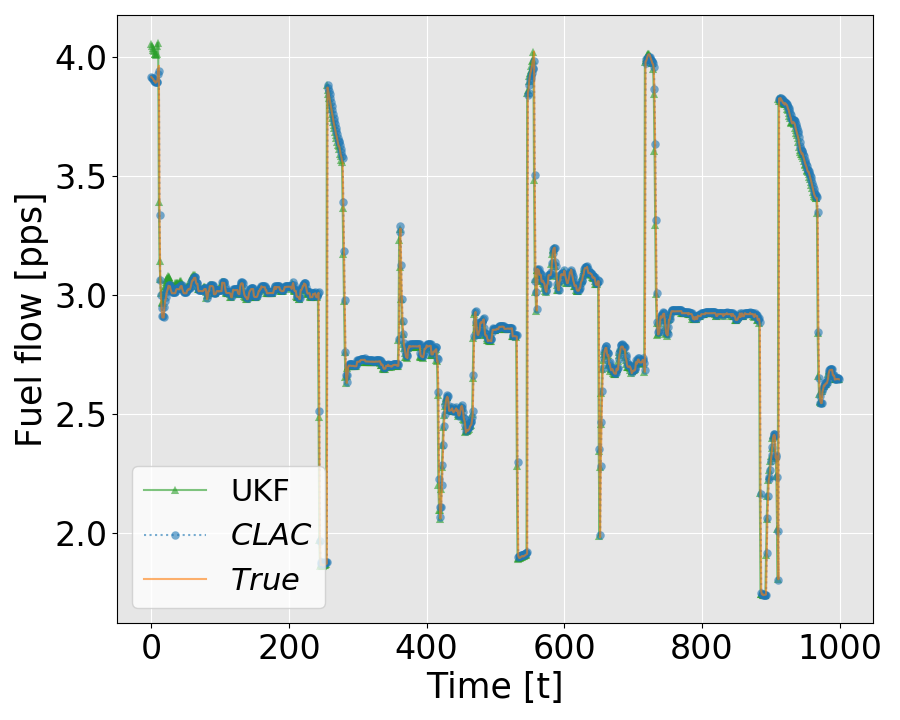}
\caption{Tracking with the proposed CLAC method (blue dots) and UKF (green triangles) on a subset of Dataset \#1 for four sensor outputs. Ground truth is shown with the orange solid line. The CLAC policy is able to keep all these sensors on track.}
\label{fig:Tracking Result}
\end{figure}

\textbf{Robustness to Sensor Noise.}
In real scenarios, the observations are always noisy. Therefore, it is also important to obtain a policy that is robust to sensor noise. To evaluate this effect, we modelled the engine sensor noise and generated a noisy dataset by adding Gaussian noise with an intensity of 70 db signal to noise ratio ($SNR_{db}=70$) to the original dataset. Table \ref{tb:sensor_noise} shows the impact of noise on the inference performance of the UKF and CLAC methods. In this case, although our policy still shows good inference ability, UKF is more robust to sensor noise.

\begin{table}[h]
\caption[Table tb:time]{Overview of the inference performance under observation noise with UKF and CLAC approaches for complete test trajectories - Dataset \#1.}
\label{tb:sensor_noise}
  \begin{center}
    \begin{tabular}{|c||c|c|} \hline
      \multicolumn{3}{|c|}{ Observation Noise: $x_s + \epsilon$ }     \\ \hline \hline
       Intensity             & UKF               &  CLAC        \\ \hline
       $\text{SNR}_{db}=70$  & \textbf{3.72e-04} &  7.18e-04    \\ \hline
    \end{tabular}
  \end{center}
\end{table}

\textbf{Tracking Accuracy.} We formulate the calibration problem as a tracking problem and use reinforcement learning to track the operational trajectories of the real systems (i.e., the observations) while being constrained to have a stable policy. Therefore, we evaluate the error between the observed real system response and the calibrated model output. Figure \ref{fig:Tracking Result} shows that our policies exhibit good tracking ability for the model outputs. Table \ref{tb:tracking} provides a complete overview of the root-mean-square error (RMSE) for each of the evaluated test cases. Although the CLAC framework shows good tracking ability in all the setups, the UKF achieves better tracking. This is an expected situation with the current RL formulation as the reinforcement learning is actually solving a more complicated problem. In particular, the current state contains the output of the DNN model $\hat{x}^{(t)}$ instead of the historical observation ($x^{(t)}$), as a result of which small errors accumulate. On the other hand, it is precisely this aspect that ensures that the proposed policy action will generalize well to unseen degradation trajectories.

\begin{table}[h]
\caption[Table tb:tracking]{Overview of the tracking performance given by RMSE with UKF and CLAC approaches in both datasets.}
  \begin{center}
    \begin{tabular}{|c||c|c|}
      \hline
       Method             & Dataset \#1     &  Dataset \#2    \\ \hline
       UKF                & \textbf{0.62}   &  \textbf{1.78}  \\ \hline
       CLAC               & 0.98            &  5.54           \\ \hline   
    \end{tabular}
  \end{center}
\label{tb:tracking}
\end{table}

\subsection{Ablation Study}

\begin{figure}[h]
\centering
\includegraphics[width=0.45\columnwidth]{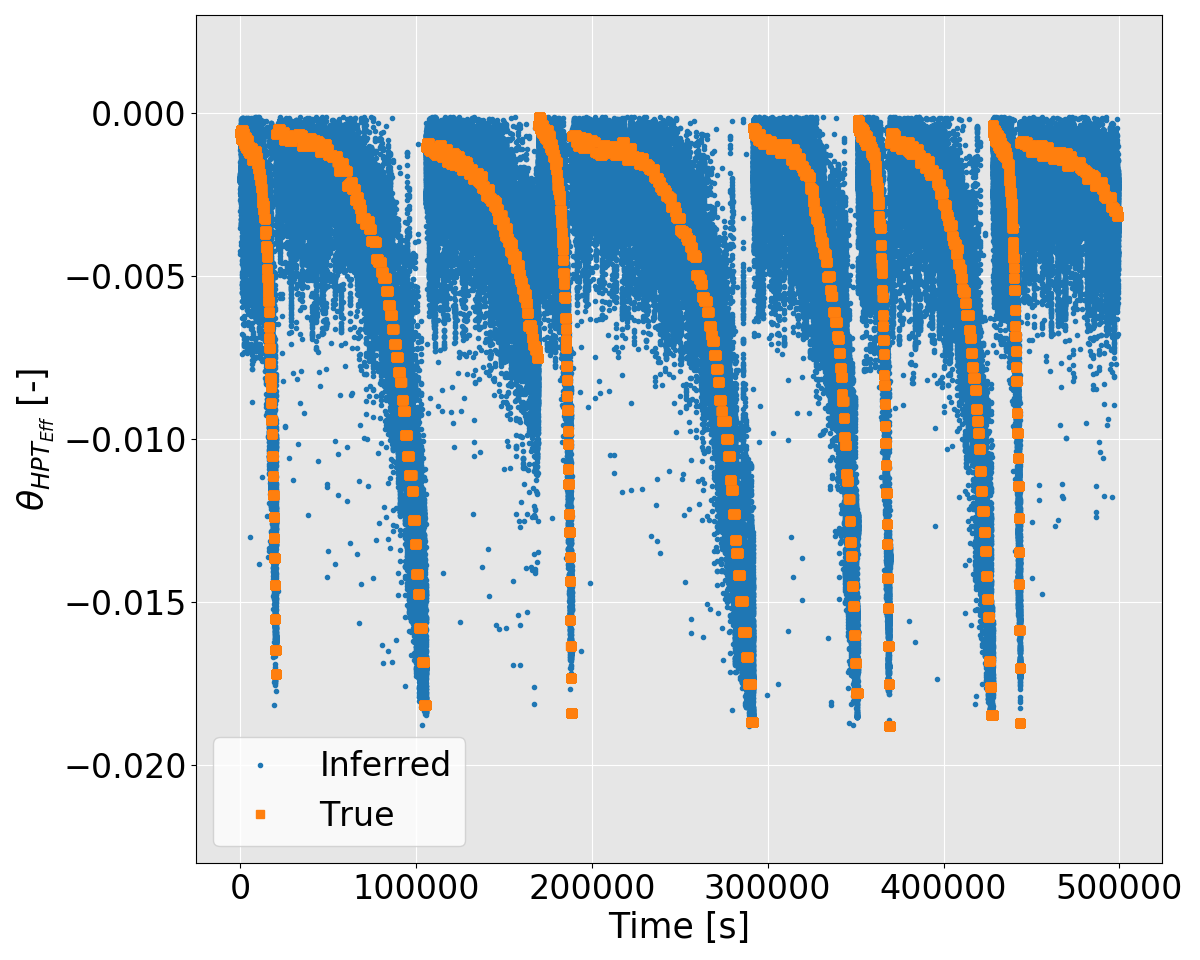}
\caption{Inferred policy's actions with LAC algorithm (blue dots) and ground truth (orange squares) in Dataset \#1. The trajectories of the ten units, stacked one after the other, are shown. SAC policy exhibits quite large variance. In contrast, as shown in Figure \ref{fig:Inference_Comp}, the CLAC policy shows a good stability and a very good match to the ground truth.}
\label{fig:RSAC}
\end{figure}

\textbf{Comparison between LAC and CLAC algorithms} 
We propose to extend LAC to CLAC to improve the stability of the policy under noisy conditions. To demonstrate the benefit of the proposed extension, we compared the inference performance of both algorithms, LAC and CLAC, on Dataset \#1. In the C-MAPSS experiments, the flying conditions are very diverse and the DNN model is not very accurate and is particularly noisy. Therefore, the DNN model may lead to an unstable policy. Figure \ref{fig:RSAC} shows the policy's actions with LAC algorithm (orange squares) and ground truth (blue dots) for the entire trajectories and demonstrates a significant reduction in the variance of the policy. Concretely, in terms of the RMSE metric, the LAC results in a RMSE of 1.3 e-3 while the CLAC led to an RMSE of 3.3e-04. Therefore, CLAC provides a $4 \times$ inference improvement.     

\section{Conclusions and future work}
We proposed a maximum entropy reinforcement learning framework and the constrained Lyapunov-based actor-critic (CLAC) algorithm for model calibration. The proposed calibration methodology achieves high inference accuracy and robustness while reducing the computational load to a level that makes the proposed methodology applicable to real-time, noisy, and large-scale calibration problems. This capability was achieved purely on the basis of training in a simulation environment without any tedious sampling or computationally expensive solution of an inverse problem. Moreover, and in contrast to the end-to-end learning architectures, the proposed methodology only requires access to the model and the observations, eliminating the need for any ground truth calibration parameters for training. Overall, the proposed CLAC algorithm achieves more precise and faster inference than the prior state-of-the-art while being more robust to system model uncertainty.

The proposed framework can be generally combined with various RL algorithms, or can even be extended to the meta RL \cite{rakelly2019efficient, finn2017model} or hierarchical RL \cite{dietterich2000hierarchical, barto2003recent}. All our experiments are currently performed in a simulated environment. As a next step, we plan to evaluate the resulting policies on the real industrial plants or robots.

Although the learning framework presented in the work is demonstrated in a model-based diagnostics task, it is applicable to any physics-based model,including those used in so-called "digital twins". Therefore, the results presented in this paper suggest a promising research direction in the field of model calibration. From an application perspective, the targeted model-based diagnostics problem was solved using exclusively a set of three deep neural networks. Therefore, the proposed framework is a paradigm shift in the field of model-based diagnostics. Starting with a model-based problem, we demonstrate that a clever arrangement of deep neural networks can learn both the relevant physics of a complex system and the inference techniques required for diagnostics. It is worth pointing out that the use of deep neural networks is very diverse (e.g., functioning as the surrogate of a physics-based system model or as an inference network in a decision-making problem). The proposed framework demonstrates the great potential of fusing physics-based and deep learning models.

\bibliographystyle{unsrt}
\bibliography{references}

\begin{thebibliography}{10}

\bibitem{Kennedy2001}
Marc~C. Kennedy and Anthony O'Hagan.
\newblock {Bayesian calibration of computer models}.
\newblock {\em Journal of the Royal Statistical Society: Series B (Statistical
  Methodology)}, 63(3):425--464, 2001.

\bibitem{Elsheikh2015}
Ahmed~H. Elsheikh, Vasily Demyanov, Reza Tavakoli, Mike~A. Christie, and
  Mary~F. Wheeler.
\newblock {Calibration of channelized subsurface flow models using nested
  sampling and soft probabilities}.
\newblock {\em Advances in Water Resources}, 75:14--30, jan 2015.

\bibitem{Sanso2008}
Bruno Sans{\'{o}}, Chris~E Forest, and Daniel Zantedeschi.
\newblock {Inferring Climate System Properties Using a Computer Model}.
\newblock Technical Report~1, 2008.

\bibitem{Henderson2009}
Daniel~A. Henderson, Richard~J. Boys, Kim~J. Krishnan, Conor Lawless, and
  Darren~J. Wilkinson.
\newblock {Bayesian emulation and calibration of a stochastic computer model of
  mitochondrial DNA deletions in substantia nigra neurons}.
\newblock {\em Journal of the American Statistical Association},
  104(485):76--87, mar 2009.

\bibitem{Rutter2009}
{Carolyn M.} Rutter, {Diana L} Miglioretti, and {James E.} Savarino.
\newblock Bayesian calibration of microsimulation models.
\newblock {\em Journal of the American Statistical Association},
  104(488):1338--1350, 12 2009.

\bibitem{Liu2019}
Shuaiqiang Liu, Anastasia Borovykh, Lech~A. Grzelak, and Cornelis~W. Oosterlee.
\newblock {A neural network-based framework for financial model calibration}.
\newblock {\em Journal of Mathematics in Industry}, 9(1):1--28, dec 2019.

\bibitem{Deng2008}
Zui~Cha Deng, Jian~Ning Yu, and Liu Yang.
\newblock {An inverse problem of determining the implied volatility in option
  pricing}.
\newblock {\em Journal of Mathematical Analysis and Applications},
  340(1):16--31, apr 2008.

\bibitem{Kangasraasio2019}
Antti Kangasr{\"{a}}{\"{a}}si{\"{o}}, Jussi P.~P. Jokinen, Antti Oulasvirta,
  Andrew Howes, and Samuel Kaski.
\newblock {Parameter Inference for Computational Cognitive Models with
  Approximate Bayesian Computation}.
\newblock {\em Cognitive Science}, 43(6), jun 2019.

\bibitem{Kumar2013}
Natarajan~Chennimalai Kumar, Arun~K. Subramaniyan, Liping Wang, and Gene Wiggs.
\newblock {Calibrating transient models with multiple responses using bayesian
  inverse techniques}.
\newblock In {\em Proceedings of the ASME Turbo Expo}, volume 7 A. American
  Society of Mechanical Engineers Digital Collection, nov 2013.

\bibitem{Higdon2008}
Dave Higdon, James Gattiker, Brian Williams, and Maria Rightley.
\newblock Computer model calibration using high-dimensional output.
\newblock {\em Journal of the American Statistical Association},
  103(482):570--583, 2008.

\bibitem{Roychoudhury2013}
I.~{Roychoudhury}, V.~{Hafiychuk}, and K.~{Goebel}.
\newblock Model-based diagnosis and prognosis of a water recycling system.
\newblock In {\em 2013 IEEE Aerospace Conference}, pages 1--9, 2013.

\bibitem{Crassidis2011}
John~L. Crassidis and John~L. Junkins.
\newblock {\em Optimal Estimation of Dynamic Systems, Second Edition (Chapman
  {\&} Hall/CRC Applied Mathematics {\&} Nonlinear Science)}.
\newblock Chapman {\&} Hall/CRC, 2nd edition, 2011.

\bibitem{Sacks1989}
J~Sacks, W~J Welch, J~S~B Mitchell, and P~W Henry.
\newblock {Design and Experiments of Computer Experiments}, 1989.

\bibitem{AriasChao2015}
Manuel {Arias Chao}, Darrel~S. Lilley, Peter Math{\'{e}}, and Volker
  Schlo{\ss}hauer.
\newblock {Calibration and Uncertainty Quantification of Gas Turbine
  Performance Models}.
\newblock In {\em Proceedings of the ASME Turbo Expo}, volume~7A, page
  V07AT29A001, 2015.

\bibitem{Kalman1960}
R.~E. Kalman.
\newblock {A new approach to linear filtering and prediction problems}.
\newblock {\em Journal of Fluids Engineering, Transactions of the ASME},
  82(1):35--45, mar 1960.

\bibitem{Einicke1999}
Garry~A. Einicke and Langford~B. White.
\newblock {Robust extended Kalman filtering}.
\newblock {\em IEEE Transactions on Signal Processing}, 47(9):2596--2599, 1999.

\bibitem{Borguet2012}
S.J Borguet.
\newblock {\em {Variations on the Kalman Filter for Enhanced Performance
  Monitoring of Gas Turbine Engines}}.
\newblock Phd thesis, Universit{\'{e}} de Li{\`{e}}ge, 2012.

\bibitem{Julier1997}
Simon~J Julier and Jeffrey~K Uhlmann.
\newblock {New extension of the Kalman filter to nonlinear systems}.
\newblock In {\em Signal Processing, Sensor Fusion, and Target Recognition VI},
  volume 3068, page 182, 1997.

\bibitem{Turner2010}
Ryan Turner and Carl~Edward Rasmussen.
\newblock {Model based learning of sigma points in unscented Kalman filtering}.
\newblock In {\em Proceedings of the 2010 IEEE International Workshop on
  Machine Learning for Signal Processing, MLSP 2010}, pages 178--183, 2010.

\bibitem{Kantas2015}
Nikolas Kantas, Arnaud Doucet, Sumeetpal~S Singh, Jan Maciejowski, and Nicolas
  Chopin.
\newblock {On Particle Methods for Parameter Estimation in State-Space Models}.
\newblock {\em Statistical Science}, 30(3):328--351, 2015.

\bibitem{Rasmussen2006}
Carl~Edward. Rasmussen and Christopher K.~I. Williams.
\newblock {\em {Gaussian processes for machine learning}}.
\newblock MIT Press, 2006.

\bibitem{damianou13a}
Andreas Damianou and Neil Lawrence.
\newblock Deep gaussian processes.
\newblock In Carlos~M. Carvalho and Pradeep Ravikumar, editors, {\em
  Proceedings of the Sixteenth International Conference on Artificial
  Intelligence and Statistics}, volume~31 of {\em Proceedings of Machine
  Learning Research}, pages 207--215, Scottsdale, Arizona, USA, 29 Apr--01 May
  2013. PMLR.

\bibitem{Marmin2018VariationalCO}
S{\'e}bastien Marmin and Maurizio Filippone.
\newblock Variational calibration of computer models.
\newblock {\em ArXiv}, abs/1810.12177, 2018.

\bibitem{Zhang2020}
Y.~{Zhang}, L.~{Guo}, B.~{Gao}, T.~{Qu}, and H.~{Chen}.
\newblock Deterministic promotion reinforcement learning applied to
  longitudinal velocity control for automated vehicles.
\newblock {\em IEEE Transactions on Vehicular Technology}, 69(1):338--348,
  2020.

\bibitem{bucsoniu2018reinforcement}
Lucian Bu{\c{s}}oniu, Tim de~Bruin, Domagoj Toli{\'c}, Jens Kober, and Ivana
  Palunko.
\newblock Reinforcement learning for control: Performance, stability, and deep
  approximators.
\newblock {\em Annual Reviews in Control}, 2018.

\bibitem{kumar2016learning}
Vikash Kumar, Abhishek Gupta, Emanuel Todorov, and Sergey Levine.
\newblock Learning dexterous manipulation policies from experience and
  imitation.
\newblock {\em arXiv preprint arXiv:1611.05095}, 2016.

\bibitem{xie2019iterative}
Zhaoming Xie, Patrick Clary, Jeremy Dao, Pedro Morais, Jonathan Hurst, and
  Michiel van~de Panne.
\newblock Iterative reinforcement learning based design of dynamic locomotion
  skills for cassie.
\newblock {\em arXiv preprint arXiv:1903.09537}, 2019.

\bibitem{hwangbo2019learning}
Jemin Hwangbo, Joonho Lee, Alexey Dosovitskiy, Dario Bellicoso, Vassilios
  Tsounis, Vladlen Koltun, and Marco Hutter.
\newblock Learning agile and dynamic motor skills for legged robots.
\newblock {\em Science Robotics}, 4(26):eaau5872, 2019.

\bibitem{sutton1992reinforcement}
Richard~S Sutton, Andrew~G Barto, and Ronald~J Williams.
\newblock Reinforcement learning is direct adaptive optimal control.
\newblock {\em IEEE Control Systems Magazine}, 12(2):19--22, 1992.

\bibitem{levine2018reinforcement}
Sergey Levine.
\newblock Reinforcement learning and control as probabilistic inference:
  Tutorial and review, 2018.

\bibitem{Todorov2008}
Emanuel Todorov.
\newblock {General duality between optimal control and estimation}.
\newblock In {\em Proceedings of the IEEE Conference on Decision and Control},
  pages 4286--4292, 2008.

\bibitem{Kappen2009}
B~Kappen, V~Gomez, and M~Opper.
\newblock {Optimal control as a graphical model inference problem}.
\newblock Technical Report arXiv:0901.0633, Jan 2009.
\newblock Comments: 12 pages.

\bibitem{Toussaint2009}
Marc Toussaint.
\newblock {Robot trajectory optimization using approximate inference}.
\newblock In {\em ACM International Conference Proceeding Series}, volume 382,
  pages 1--8, New York, New York, USA, 2009. ACM Press.

\bibitem{Ziebart2010}
Brian~D. Ziebart, J.~Andrew Bagnell, and Anind~K. Dey.
\newblock Modeling interaction via the principle of maximum causal entropy.
\newblock In {\em Proceedings of the 27th International Conference on
  International Conference on Machine Learning}, ICML’10, page 1255–1262,
  Madison, WI, USA, 2010. Omnipress.

\bibitem{sutton1998introduction}
Richard~S Sutton, Andrew~G Barto, et~al.
\newblock {\em Introduction to reinforcement learning}, volume 135.
\newblock MIT press Cambridge, 1998.

\bibitem{ziebart2010modeling}
Brian~D Ziebart.
\newblock Modeling purposeful adaptive behavior with the principle of maximum
  causal entropy.
\newblock 2010.

\bibitem{haarnoja2018soft}
Tuomas Haarnoja, Aurick Zhou, Kristian Hartikainen, George Tucker, Sehoon Ha,
  Jie Tan, Vikash Kumar, Henry Zhu, Abhishek Gupta, Pieter Abbeel, et~al.
\newblock Soft actor-critic algorithms and applications.
\newblock {\em arXiv preprint arXiv:1812.05905}, 2018.

\bibitem{tian2019model}
Yuan Tian.
\newblock Model free reinforcement learning with stability guarantee.
\newblock 2019.

\bibitem{Ljung1990}
Lennart Ljung and Svante Gunnarsson.
\newblock {Adaptation and tracking in system identification-A survey}.
\newblock {\em Automatica}, 26(1):7--21, jan 1990.

\bibitem{mnih2015human}
Volodymyr Mnih, Koray Kavukcuoglu, David Silver, Andrei~A Rusu, Joel Veness,
  Marc~G Bellemare, Alex Graves, Martin Riedmiller, Andreas~K Fidjeland, Georg
  Ostrovski, et~al.
\newblock Human-level control through deep reinforcement learning.
\newblock {\em Nature}, 518(7540):529--533, 2015.

\bibitem{lillicrap2015continuous}
Timothy~P Lillicrap, Jonathan~J Hunt, Alexander Pritzel, Nicolas Heess, Tom
  Erez, Yuval Tassa, David Silver, and Daan Wierstra.
\newblock Continuous control with deep reinforcement learning.
\newblock {\em arXiv preprint arXiv:1509.02971}, 2015.

\bibitem{haarnoja2018soft2}
Tuomas Haarnoja, Aurick Zhou, Kristian Hartikainen, George Tucker, Sehoon Ha,
  Jie Tan, Vikash Kumar, Henry Zhu, Abhishek Gupta, Pieter Abbeel, et~al.
\newblock Soft actor-critic algorithms and applications.
\newblock {\em arXiv preprint arXiv:1812.05905}, 2018.

\bibitem{Li2002}
Y.~G. Li.
\newblock {Performance-analysis-based gas turbine diagnostics: A review}.
\newblock {\em Proceedings of the Institution of Mechanical Engineers, Part A:
  Journal of Power and Energy}, 216(5):363--377, jan 2002.

\bibitem{Fentaye2019}
Fentaye, Baheta, Gilani, and Kyprianidis.
\newblock {A Review on Gas Turbine Gas-Path Diagnostics: State-of-the-Art
  Methods, Challenges and Opportunities}.
\newblock {\em Aerospace}, 6(7):83, jul 2019.

\bibitem{Urban1973}
Louis~A. Urban.
\newblock {Gas Path Analysis Applied to Turbine Engine Condition Monitoring}.
\newblock {\em Journal of Aircraft}, 10(7):400--406, jul 1973.

\bibitem{Frederick2007}
Dean~K Frederick, Jonathan~A Decastro, and Jonathan~S Litt.
\newblock {User's Guide for the Commercial Modular Aero-Propulsion System
  Simulation (C-MAPSS)}.
\newblock Technical report, 2007.

\bibitem{DASHlink}
{NASA}.
\newblock {DASHlink - Flight Data For Tail 687}, 2012.

\bibitem{AriasChao2020}
Manuel {Arias Chao}, Chetan~S Kulkarni, Kai Goebel, and Olga Fink.
\newblock Damage propagation modeling for aircraft engine run-to-failure
  simulation under real flight conditions.
\newblock {\em Under review}, 2020.

\bibitem{Chapman2017}
Jeffryes~W. Chapman and Jonathan~S. Litt.
\newblock {Control Design for an Advanced Geared Turbofan Engine}.
\newblock In {\em 53rd AIAA/SAE/ASEE Joint Propulsion Conference}, Reston,
  Virginia, jul 2017. American Institute of Aeronautics and Astronautics.

\bibitem{rakelly2019efficient}
Kate Rakelly, Aurick Zhou, Deirdre Quillen, Chelsea Finn, and Sergey Levine.
\newblock Efficient off-policy meta-reinforcement learning via probabilistic
  context variables.
\newblock {\em arXiv preprint arXiv:1903.08254}, 2019.

\bibitem{finn2017model}
Chelsea Finn, Pieter Abbeel, and Sergey Levine.
\newblock Model-agnostic meta-learning for fast adaptation of deep networks.
\newblock In {\em Proceedings of the 34th International Conference on Machine
  Learning-Volume 70}, pages 1126--1135. JMLR. org, 2017.

\bibitem{dietterich2000hierarchical}
Thomas~G Dietterich.
\newblock Hierarchical reinforcement learning with the maxq value function
  decomposition.
\newblock {\em Journal of artificial intelligence research}, 13:227--303, 2000.

\bibitem{barto2003recent}
Andrew~G Barto and Sridhar Mahadevan.
\newblock Recent advances in hierarchical reinforcement learning.
\newblock {\em Discrete event dynamic systems}, 13(1-2):41--77, 2003.

\bibitem{maas2013rectifier}
Andrew~L Maas, Awni~Y Hannun, and Andrew~Y Ng.
\newblock Rectifier nonlinearities improve neural network acoustic models.
\newblock In {\em Proc. icml}, volume~30, page~3, 2013.

\bibitem{Kingma2014Adam}
Diederik~P Kingma and Jimmy~Lei Ba.
\newblock {Adam: A method for stochastic optimization}.
\newblock In {\em 3rd International Conference on Learning Representations,
  ICLR 2015 - Conference Track Proceedings}, 2015.

\bibitem{Glorot}
Xavier Glorot and Yoshua Bengio.
\newblock {Understanding the difficulty of training deep feedforward neural
  networks}.
\newblock Technical report, 2010.

\end{thebibliography}

\appendix
\section{Neural Network Architectures and Hyper-parameters}
\subsection{Reinforcement learning}
\begin{figure}[ht]
\centering
\includegraphics[width=8.5cm]{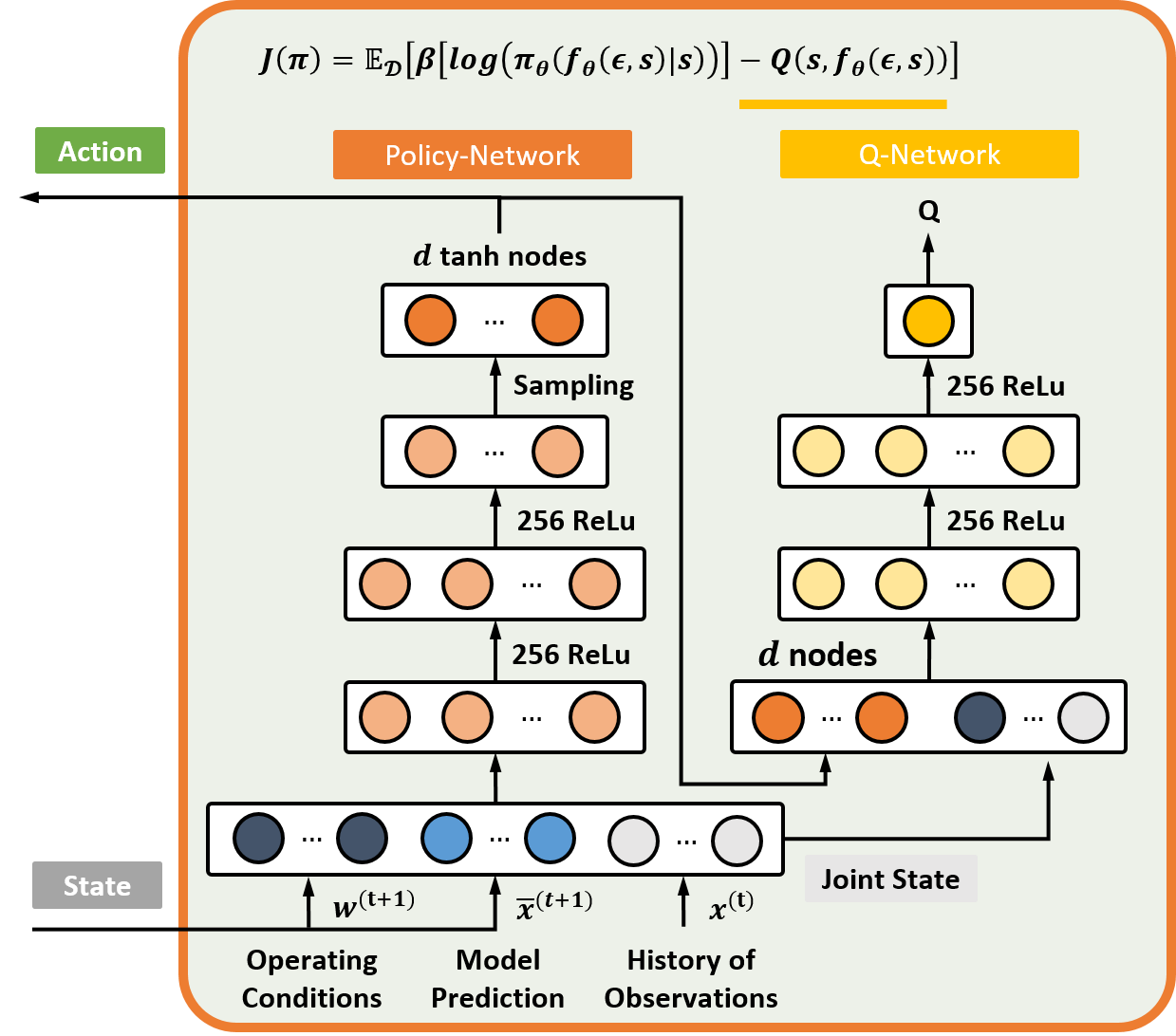}
\caption{The overview network architecture of the reinforcement learning calibration framework with CLAC. The agent observe the current state, which is described as $s_t=[\hat{x}^{(t)},x^{(t-1)},\omega^{(t)}]$. Then, the agent make a decision about action $a^{(t)}=\theta^{(t)}$ according to the observation.}
\label{fig:networks}
\end{figure}

The proposed framework and method requires three neural networks: Policy, Lyapunov and Dynamical Model networks. The overall network structure of the proposed method is shown in Figure \ref{fig:networks}.

\textbf{Policy and Lyapunov Networks}. For the policy network, we use a fully-connected multi-layer perceptron (MLP) with two hidden layers of 256 units, outputting the mean and standard deviations of a Gaussian distribution. We adopt the invertible squashing function technique as proposed in \cite{haarnoja2018soft2} to the output layer of the policy network. For the Lyapunov network, we use a fully-connected MLP with two hidden layers of 256 units, outputting the Lyapunov value. All the hidden layers use leaky-ReLU \cite{maas2013rectifier} activation function.

\textbf{Simulator Network}. The system dynamics is approximated with an MLP with four layers ($L=4$). The hidden layers have 100 units ($n^{1}=n^{2}=n^{3}=100$). The output layer has the dimension of the sensor reading vector (i.e. $n^{L}=n$). \textit{ReLU} activation function was used throughout the hidden layers. For the output layer $\sigma^L=I$ is the identity.


The optimization of the networks' weights was carried out with mini-batch stochastic gradient descent (SGD) and with the \textit{Adam} algorithm \cite{Kingma2014Adam}. \textit{Xavier} initializer \cite{Glorot} was used for the weight initializations. Table \ref{tb:settings_CLAC} provides a detailed overview of the hyperparameters used for the experiments.

\begin{table}[h]
\begin{center}
\caption{LAC and CLAC Hyperparameters}
\begin{tabular}{l|c}
Hyperparameters                   & Value   \\ \hline
Minibatch size                    & 256     \\
Learning rate - Actor             & 1e-4    \\
Learning rate - Critic            & 3e-4    \\
Learning rate - E2E               & 1e-4    \\
Target entropy                    &-d \\
Target smoothing coefficient($\tau$) & 0.005\\
Discount($\gamma$)                & 0.99    \\
$\alpha_3$                        & 1       \\
Initial $\beta$                   & 2       \\
$\lambda$                         & 0.1     \\
\label{tb:settings_CLAC}
\end{tabular}
\end{center}
\end{table}

\subsection{E2E and UKF}
\textbf{E2E Network}. To evaluate the different calibration methods under equivalent models, the E2E network is as also a MLP. In this way, we separate the effect of regularization in the form of model and learning strategies choice from other inductive bias in the form of choice of neural network type. The hidden layers have 100 units ($n^{1}=n^{2}=n^{3}=100$). The output layer has the dimension of the sensor reading vector (i.e. $n^{L}=n$). \textit{ReLU} activation function was used throughout the hidden layers. For the output layer $\sigma^L=I$ is the identity. The resulting architecture is the result of a grid reach. 

\textbf{Set-up of the UKF algorithm}
The UKF algorithm required the definition of the diagonal covariance matrices $Q$ and $R$. We assumed the covariance matrices to be diagonal matrices with normalized standard deviation $r=0.01$ and $q=0.01$ (i.e., $Q=q^2 \mathbf{I_n}$ and $R=r^2 \mathbf{I_d}$ where $\mathbf{I_k}$ is the identity matrix of dimension $k$).

\end{document}